\documentclass[twocolumn,showpacs,preprintnumbers,amsmath,amssymb]{revtex4}

 \def\XXint#1#2#3{{\setbox0=\hbox{$#1{#2#3}{\int}$}
     \vcenter{\hbox{$#2#3$}}\kern-.5\wd0}}

\def\fakebold#1{\relax\ifvmode\leavevmode\fi%
\ifmmode%
\setbox0=\hbox{$#1$}%
\else%
\setbox0=\hbox{#1}%
\fi%
\kern-.02em\copy0 \kern-\wd0%
\kern .04em\copy0 \kern-\wd0%
\kern-.0125em\raise.02em\box0%
}%

\usepackage{graphicx} \usepackage{dcolumn} \usepackage{bm}

\begin{document}



\title{Quasiparticle effective mass divergence in two dimensional
  electron systems}

\author{Ying Zhang} \author{S. Das Sarma} \affiliation{Condensed
  Matter Theory Center, Department of Physics, University of Maryland,
  College Park, MD 20742-4111}

\date{\today}

\begin{abstract}
Within the infinite series of ring (or bubble) diagram approximation
for the electronic self-energy as appropriate for the long-range
Coulomb interaction, we calculate the density-dependent $T=0$ Fermi
liquid quasiparticle effective mass renormalization in two dimensional
(2D) electron systems in the random-phase-approximation (RPA). We find
that the quasiparticle effective mass increases very strongly with
increasing $r_s$, exhibiting an unexpected divergence at a critical
$r_s$ value ($\sim 16$ in the ideal 2D system). We discuss the
possibility for and the consequences of such an interaction induced
effective mass divergence as the electron density decreases, and
critically comment on the likely theoretical scenarios which could
lead to such a low density effective mass divergence in electron
liquids.
\end{abstract}


\maketitle


\section{Introduction}
\label{sec:intro}

An interacting homogeneous quantum electron system in a positive
jellium background is characterized (at $T=0$) by the interaction
parameter $r_s$, the dimensionless average separation between
electrons measured in the unit of effective Bohr radius, with large
(small) $r_s$ being the strongly (weakly) interacting regime
corresponding respectively to low (high) densities. The interaction
parameter $r_s$ depends only on the density ($n$) of the electron
system with $r_s \propto n^{-1/3} (n^{-1/2})$ in 3D (2D) systems. The
$r_s$-parameter is essentially proportional to the dimensionless ratio
of the (interacting) potential energy to the (noninteracting) kinetic
energy, with the potential (kinetic) energy varying as $r_s^{-1}
(r_s^{-2})$. (This simple one parameter picture of interacting
electron systems is valid only at $T=0$ and for clean free-electron
jellium systems with little disorder.)

Alkali metals (e.g. Li, Na, K, Rb, Cs) are examples of interacting
three dimensional (3D) electron systems with reasonably simple band
structures so that the homogeneous free-electron jellium model may be
applicable. The typical $r_s$-parameter for 3D metals is in the range
of $3-6$ (e.g. $r_s$ = 3.2 [Li], 4.0 [Na], 5.0 [K], 5.2 [Rb], 5.6
[Cs]), making metals strongly interacting electron systems since $r_s
> 1$. Electrons (or holes, as in p-GaAs systems) confined by external
electric fields near interfaces in semiconductor structures (e.g. Si
MOSFETs, GaAs quantum wells and heterojunctions) are examples of 2D
electron systems, which are actually quite ideal free electron jellium
systems if the impurity disorder is not too high since the
noninteracting energy has free electron parabolic form without any
band structure complications. One particular advantage of
semiconductor-based 2D electron systems is that often (as in Si
MOSFETs and GaAs HIGFETs) the carrier density can be tuned over one or
two orders of magnitude (by changing the external gate voltage), thus
enabling the tuning of the $r_s$-parameter in the same sample over one
order of magnitude -- for example, in high quality Si MOSFETs, $r_s$
could be tuned from 1 to 12 and in gated n-GaAs 2D HIGFET structures,
where extremely low carrier densities can be achieved and $r_s$ can be
made as high as 15. In hole-doped 2D p-GaAs structures it is possible
to achieve extremely high values of $r_s$ ($\sim 20 - 30$), but there
are considerable band structure complications in 2D p-GaAs hole
systems due to the presence of strong spin-orbit coupling and the
complex multi-band structure of the GaAs valence band so that 2D
p-GaAs systems may not be thought of as simple isotropic
free-electron-like systems. This capability of tuning the $r_s$
parameter of a single system over a large range has made
semiconductor-based 2D electron systems attractive systems for
studying Coulomb quantum many-body effects, and indeed many-body
renormalization has been actively studied in 2D electron systems over
the last thirty years both experimentally~\cite{ando, smith,
  coleridge, pan, kravchenko1, kravchenko2, pudalov, kravchenko3,
  vitkalov} and theoretically~\cite{janak, vinter, ting, jalabert,
  marmorkos, giuliani1, giuliani2, short, chubukov1, dolgopolov,
  khodel, morawetz, jonson}.

In this paper we study theoretically the quasiparticle effective mass
($m^*$) renormalization in an interacting 2D electron system at $T=0$
finding an interesting (and intriguing) divergence in the calculated
renormalized effective mass at low (high) carrier densities ($r_s$).
Our result takes on particular significance since such an effective
mass divergence has recently been claimed to have been experimentally
observed in 2D Si MOSFETs with $r_s$ values close to (around $r_s \sim
10 - 17$)~\cite{kravchenko1, kravchenko2, kravchenko3} where we find
our theoretical divergence. Because of the necessarily approximate
(and somewhat uncontrolled) nature of our theory we cannot be
absolutely certain that the observed effective mass divergence in 2D
semiconductor systems is indeed the same phenomenon as what we obtain
in our theory, but it seems likely that the observed divergence is
related to our theoretical findings because our calculated functional
form of $m^*(r_s)$ is very similar to the experimental observation. We
emphasize that the self-energy approximation (the so-called `GW' or
the single-loop calculation) used in our theory has been extensively
used in the literature up to $r_s \approx 6-7$, and our main
contribution is to extend the calculation to even larger values of
$r_s$, mainly motivated by the recent experimental claims we mentioned
above.

The existing paradigm (and one of the key cornerstones of condensed
matter physics) for interacting electron systems is Landau's Fermi
liquid theory developed fifty years ago. The Fermi liquid theory
asserts a one-to-one low energy correspondence between the
noninteracting Fermi gas and the interacting Fermi liquid with weakly
interacting quasiparticles (with renormalized single particle
properties such as effective mass, specific heat, magnetic
susceptibility, compressibility, etc.) in the interacting system
behaving similar to free electrons in the noninteracting system. The
key idea is that the Fermi surface, i.e. the $T=0$ discontinuity in
the single-particle momentum distribution function at Fermi
wave-vector $k=k_F$, survives the effect of interaction (albeit with a
reduced or renormalized discontinuity at $k_F$) leading to
qualitatively similar leading-order low temperature behavior in the
single particle thermodynamic properties such as specific heat, spin
susceptibility and compressibility, for interacting and noninteracting
systems. The Fermi liquid theory provides a natural explanation for
the remarkable success of the noninteracting Fermi gas theory in
describing thermodynamic, magnetic and even transport properties of
metals, which are in fact strongly interacting Fermi liquids ($r_s
\approx 3 - 6$).  The question, however, naturally arises whether the
Fermi liquid theory remains valid as the interaction strength is
increased by going to higher values of $r_s$. This question has taken
on significance in view of recent experiments in 2D electron systems
at large $r_s$ values, where the measured quasiparticle effective mass
is seen to increase very rapidly with increasing
$r_s$~\cite{kravchenko1, kravchenko2, kravchenko3, pudalov}. This is
the motivation for our theoretical investigation of the quasiparticle
effective mass in 2D electron systems as a function of $r_s$ going to
low carrier densities (i.e. high $r_s$).

There are several possible theoretical scenarios for the breakdown of
the Fermi liquid theory in a clean interacting electron liquid at low
densities (or high $r_s$). At very large $r_s$ the system could
obviously reduce its potential energy (at the cost of higher kinetic
energy) by becoming an electron crystal (the so-called quantum Wigner
crystal~\cite{wigner}). The $T=0$ transition from an electron liquid
to a Wigner solid should be a first order transition in a clean
system, occurring at $r_s \sim 35-40$ for 2D systems as deduced from
quantum Monte Carlo calculations~\cite{tanatar}. The other scenario is
the possibility of an interaction driven phase transition to some
other ordered phase (e.g. charge density wave, spin density wave,
superconductor, ferromagnet). We do not consider the possibility of
such interaction driven phase transitions in the current work,
assuming that the system remains a homogeneous electron liquid for all
values of $r_s$ we investigate. Our goal is to calculate the 2D zero
temperature quasiparticle effective mass $m^*(r_s)$ as a function of
the density parameter $r_s$ in the interacting electron liquid.

One problem, which has no rigorous solution, that immediately arises
in this context is that the electron system for $r_s > 1$ is by
definition a strongly interacting system with a formal failure of the
usual Feynman-Dyson diagrammatic perturbation theory, and in fact
there is no known controlled technique for systematically calculating
exact quasiparticle properties in interacting electron systems for
$r_s > 1$. In the weak coupling limit, $r_s \ll 1$, it has been known
for a long time that an asymptotically exact many-body perturbation
theory can be developed for quasiparticle properties by expanding in
the $r_s$ parameter. In fact, in the $r_s \ll 1$ limit, the expansion
of the electron self-energy (as well as the polarizability) in the
infinite series of ring or bubble diagrams (the so-called `RPA', the
random-phase-approximation) is exact because of the long-range nature
of Coulomb interaction. Thus in the $r_s \to 0$ limit it is possible
to obtain exact results for the effective mass renormalization in the
interacting electron system (interacting via the Coulomb interaction).
Such an exact result is of course useless for real interacting
electron systems (e.g. metals in 3D, semiconductor layers in 2D) which
are never in the weakly interacting $r_s \to 0$ limit, but are
actually in the strongly interacting $r_s > 1$ regime. Therefore,
analytical results obtained by expanding in $r_s$ (assumed to be a
small parameter) are completely irrelevant for any practical purpose
as was already emphasized by Hedin~\cite{hedin} and by
Rice~\cite{rice} a long time ago, since real 2D and 3D electron
systems are always in the $r_s > 1$ regime. There is actually a deeper
(and somewhat subtle) problem with the $r_s$ expansion of quasiparticle
properties in the $r_s \to 0$ limit. It has been known for a
while~\cite{luttinger} that the $r_s$-expansion for the interacting
electron liquid problem formally fails at $T=0$ even in the
high-density $r_s \to 0$ limit since the radius of convergence for
such a power series expansion in $r_s$ is, strictly speaking, zero. In
fact, a weakly interacting ($r_s \ll 1$) system of fermions does not
remain a normal Fermi liquid down to zero temperature, but makes a
transition to a superconducting state (arising purely out of the
repulsive electron-electron interaction) in some rather high orbital
angular momentum channel with an exponentially small value of $T_c$.
In our discussion we assume that this so-called Kohn-Luttinger
superconductivity can be ignored since $T_c$ is exponentially small
for this transition. We uncritically assume the existence of a normal
Fermi liquid for the purpose of this work.

We therefore need a {\em robust} (but necessarily inexact except in
the $r_s \to 0$ limit) many-body approximation for calculating the
quasiparticle effective mass at arbitrary values of $r_s$. Following
Hedin and Rice~\cite{hedin, rice} we use the RPA as our central
approximation although we improve upon RPA in some situations by using
various approximation schemes as described in details later in this
paper. We do want to make an important point here about the validity
(or lack thereof) of RPA in the strongly interacting $r_s > 1$ regime.
It is often stated or assumed that RPA is invalid for $r_s > 1$. This
assertion is false as has already been argued~\cite{hedin, rice,
  dubois} a long time ago in the context of the applicability of RPA
to 3D metals ($r_s \approx 3 - 6$). For arbitrary values of $r_s$, it
is more appropriate to think of RPA as a dynamical Hatree-Fock
self-energy calculation (using the dynamically screened interaction)
rather than a perturbation expansion in $r_s$ (which makes sense only
in the $r_s \to 0$ high density asymptotic limit).  Such a dynamical
mean-field calculation (called the `GW' approximation~\cite{hedin},
where G is the electron Green function and W the dynamically screened
Coulomb interaction) has a long history in electronic systems, and is
extensively employed in quasiparticle band structure calculations
(going beyond LDA) in 3D systems (metals, insulators, and
semiconductors) with very impressive empirical success. The RPA is a
systematic expansion in the dynamically screened interaction (i.e. an
infinite resummation in terms of the bare unscreened Coulomb
interaction which has the long-range divergence making an expansion in
bare interaction, even on a formal level, impossible), and in general,
the validity of RPA for arbitrary $r_s$ is not rigorously known except
that it is rigorously known to be {\em exact} in the $r_s \to 0$
limit. The reverse is, however, not true, i.e., RPA is not necessarily
a bad approximation for $r_s > 1$ (where one needs to do the
calculations numerically, of course, as we do in this work instead of
carrying out an $r_s$ expansion). The applicability of RPA for $r_s >
1$ (through exact numerical calculations without any $r_s$ expansion)
must be ascertained empirically, and in view of its widespread
empirical success in both 3D (metals) and 2D (semiconductor space
charge layers) electron systems in the $r_s > 1$ regime, we have
decided to carry out an RPA-based quasiparticle effective mass
calculation in 2D systems for larger $r_s$ values in the regime of
current interest in 2D systems~\cite{kravchenko1, kravchenko2,
  pudalov, kravchenko3, vitkalov}.  We do provide some qualitative and
quantitative justification for the possible validity of RPA
self-energy calculations at large $r_s$ in section~\ref{sec:RPA} of
this paper. We argue that our RPA calculation of $m^*(r_s)$ giving a
divergent effective mass at a large $r_s$ value ($\sim 16$ in the
ideal 2D system) is qualitatively valid although the precise critical
$r_s$ value where the divergence occurs may not be quantitatively
accurate within RPA. Our conclusion about the qualitative validity of
the RPA prediction of a divergent 2D effective mass at large $r_s$ is
further strengthened by our finding that the inclusion of higher order
vertex corrections in the theory (both in the polarizability and in
the self-energy in a consistent manner) does {\em not} change our
qualitative conclusion at all.

We believe that the terminology `RPA' is a misnomer for the
leading-order dynamical screening self-energy calculation we are
carrying out in this paper although this has now become the standard
terminology in the literature. A more appropriate name for this
particular approximation scheme (often also called the `GW
approximation' in the context of electronic structure calculation) is
the `Ring Diagram Approximation' or `RDA', which emphasizes the basic
physics of keeping only the infinite series of ring or bubble diagrams
in calculating the self-energy. The two most crucial non-perturbative
features of the interacting electron physics are preserved (in fact,
emphasized) in this `RDA': Screening of the long-range interaction to
eliminate the Coulomb divergence in each order and the collective
plasmon excitation of the interacting system. Inclusion of these two
key {\em non-perturbative} aspects of the physics of the problem makes
RPA (i.e. `RDA') a reasonable approximation for arbitrary values of
$r_s$ as long as there is no quantum phase transition to a new ground
state.

It is well-known~\cite{book} that the in $r_s \to 0$ extreme high
density limit, the RPA calculation for the quasiparticle effective
mass becomes exact giving the following asymptotic formula (in both 2D
and 3D)
\begin{equation}
\label{eq:mlrs}
\left. {m^* \over m} \right|_{r_s \to 0} = 1 + a r_s (b + \ln r_s) +
{\cal O}(r_s^2),
\end{equation}
where $a$, $b$ are constants of order unity in both 2D and 3D. This
formula shows that $m^*(r_s)$ has a nontrivial $r_s$ expansion even in
the $r_s \to 0$ limit where the ring diagrams of RPA asymptotically
dominate the self-energy. We emphasize that $m^*(r_s) < m$, i.e. the
mass renormalization is negative for small $r_s$ according to the
high-density expansion. Our goal in this paper is to obtain the RPA
effective mass numerically without resorting to any expansion -- the
only approximation we utilize is the expansion in effective
interaction, which we will argue later in section \ref{sec:RPA} is
{\em not} an expansion in $r_s$ but in some other effective parameter
$\delta$ (a function of $r_s$) that may actually be small even for
$r_s$ larger than unity. We will argue later in this paper that the
actual expansion parameter in our theory is not the physical density
parameter $r_s$, but a more subtle parameter $\delta= r_s (1+ c
r_s)^{-1}$ where $c \gg 1$ making the theory exact in the high-density
$\delta \to 0$ (i.e. $r_s \to 0$), but reasonably accurate for $r_s >
1$ also.

As mentioned above, the low-density ($r_s > 5$) Fermi liquid behavior
of 2D electron systems has taken on special significance during the
last few years due to the extensive experimental
activity~\cite{kravchenko1, kravchenko2, pudalov, kravchenko3,
  vitkalov} in the subject generically referred to as the
two-dimensional metal-insulator-transition (2D MIT), where lowering
density in clean 2D electron systems causes an apparently sharp
localization transition in the system. The critical density for the 2D
MIT seems to be rather close (perhaps even the same) as the density
for the apparent effective mass divergence~\cite{kravchenko1,
  kravchenko2}. It may appear trivially obvious that the conductivity
$\sigma = n e^2 \tau /m^*$, where $n / \tau / m^*$ are the carrier
density / transport relaxation time / effective mass respectively,
will vanish if the effective mass $m^*$ diverges (therefore causing an
MIT at the critical $r_s$ value where $m^*$ diverges), but this is not
necessarily true. In particular, in an ideal system (i.e. a
translationally invariant disorder-free jellium-background electron
liquid) the effective mass entering the conductivity formula is the
bare (band) effective mass (and {\em not} the renormalized
quasiparticle effective mass), and therefore a strong increase in the
renormalized quasiparticle mass $m^*$, as we obtain in this work,
should not affect transport properties in a translationally invariant
electron liquid. The reason for this is very simple: In a
translationally invariant system the total Hamiltonian can be divided
into two parts, a center of mass Hamiltonian and a relative coordinate
Hamiltonian, with electron-electron interactions affecting only the
relative coordinate part which does not affect long wavelength
transport by virtue of Galilean invariance. Real 2D systems, even the
very high-quality (i.e. low disorder) systems where the 2D MIT
phenomena are typically observed, are not ideal and have considerable
impurity disorder in them. It is, therefore, possible that the
renormalized quasiparticle effective mass appears in the formula for
the conductivity of real 2D electron systems, with the 2D metal to
insulator transition at low carrier densities somehow being
intrinsically connected with the interaction induced low density
effective mass divergence phenomenon since translational invariance
does not apply to real systems. But the issue of the relevant
effective mass entering the transport theory (e.g. the low temperature
conductivity) in a disordered system is extremely subtle, and it is by
no means clear that the many-body quasiparticle effective mass
$m^*(r_s)$ could be used in obtaining the conductivity. Much more work
is needed to clarify this issue.

Our finding of a divergent quasiparticle effective mass in the
strongly interacting 2D electron system is very similar to the
corresponding situation~\cite{casey, boronat} in normal He-3 which is
known to be an ``almost localized'' Fermi liquid ~\cite{vollhardt}
with a divergent effective mass arising from the strong inter-particle
interaction (which is short-ranged in He-3). Similar to the situation
in 2D electron systems~\cite{kravchenko3, vitkalov, pudalov}, He-3
exhibits a strongly enhanced, almost ferromagnetic, susceptibility. We
believe that this similarity in the quasiparticle properties of normal
He-3 and low-density 2D electron gas is not fortuitous, and ``almost
localization'' (i.e. very heavy or divergent effective mass) and
``almost ferromagnetism'' (i.e. strongly enhanced susceptibility) may
actually be the generic behavior of strongly interacting Fermi
liquids, provided the interaction is sufficiently strong, independent
of whether the interaction is long-ranged (as in our case) or
short-ranged (as in He-3). The presence of disorder in 2D electron
systems may, however, add considerable complications to the 2D
semiconductor systems not present in He-3. It is interesting to note
that the strong-coupling regime (with large $m^*/m$) is attained in
the electron system at {\em low} densities whereas the strong-coupling
regime with large $m^*/m$ is achieved at {\em high} densities in He-3
with the difference arising from the long-range (short-range) nature
of the interaction in electrons (He-3).

There have of course been many other calculations of $m^*(r_s)$ in
both 3D and 2D electron systems using the RPA self-energy
approximation. None of these existing calculations has however been
extended to large $r_s$ values (i.e. low density), and therefore the
actual divergence of the effective mass $m^*(r_s)$ at a critical $r_s
(\sim 16.6)$ we find in this work has not been discussed before in the
literature. Motivated by the 2D MIT physics, however, there have been
several recent theoretical discussions~\cite{dolgopolov, khodel,
  morawetz} of a divergent 2D effective mass at low electron-density,
but the underlying physics of these works~\cite{dolgopolov, khodel,
  morawetz} are different from the strong interaction physics
discussed in our work. Although we believe (and argue in this paper)
that RPA remains (at least qualitatively) a good approximation even at
low carrier densities, the question remains whether the divergence of
$m^*(r_s)$ we discover within the RPA self-energy calculation is real
(i.e. qualitatively valid) or an artifact. The related question is
whether the Fermi liquid theory applies near the effective mass
divergence, or equivalently, what the nature of this mass divergence
behavior is on a fundamental level (i.e. is it some kind of a quantum
phase transition, and if it is, what is the nature of the quantum
phase on the other side of the transition beyond the mass
divergence?). Unfortunately we are not in a position to definitively
answer these questions. But we could certainly speculate on the
various theoretical possibilities.  First, we believe our RPA
calculation to be qualitatively valid, i.e.  we believe that
$m^*(r_s)$ does indeed increase very strongly at low densities,
possibly diverging at a critical density. Second, the divergent
effective mass obviously signifies some kind of a strong-coupling
transition whose nature cannot be surmised on the basis of our Fermi
liquid calculations since the basic premise of the one-to-one
correspondence with the noninteracting electron gas breaks down at the
transition point.  These are all important and interesting questions
which should be further investigated in future work. In this paper, we
only report the finding of a low-density quasiparticle effective mass
divergence in 2D electron liquids within the RPA self-energy
calculation.

In section~\ref{sec:form} we present the formalism of our theoretical
approach. In section~\ref{sec:results} we provide our RPA results as a
function of density. In section~\ref{sec:RPA} we discuss going beyond
RPA in the process providing theoretical justification for our
calculation. We conclude in section~\ref{sec:con} with a discussion.


\section{Formalism}
\label{sec:form}

We describe the basic RPA self-energy formalism in \ref{sec:form:E},
followed by the discussion of the finite quasi-two dimensional layer
width correction in \ref{sec:form:W}. We discuss appropriate inclusion
of higher-order vertex corrections (beyond RPA) in \ref{sec:form:V}
(more on this in section V), and the on-shell or off-shell calculation
of the quasiparticle effective mass in \ref{sec:form:m}.

\subsection{RPA self energy}
\label{sec:form:E}

\begin{figure}[htbp]
\centering \includegraphics[width=2in]{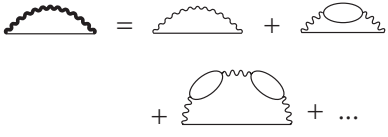}
  \caption{Feynman diagram for self-energy in the RPA
    calculation. Solid lines denote the free electron Green's function
    $G_0$ and the wiggly lines the bare Coulomb potential. Note that
    formally the RPA self-energy is $G_0 u$, where $u$ is the
    dynamically screened interaction in the ring diagram
    approximation.}
\label{fig:feynman}
\end{figure}

RPA effective mass calculation simply means that the quasiparticle
self-energy is obtained using RPA dynamical screening approximation.
Fig.~\ref{fig:feynman} shows the Feynman diagram for RPA self-energy,
from which we obtain (see, e.g., \cite{book}) for the 2D self-energy
($\hbar = 1$)
\begin{equation}
\label{eq:E0}
\Sigma({\bf k}, \omega) = - \int {d^2 q \over (2 \pi)^2} 
\int {d \nu \over 2 \pi i} {v_q \over \epsilon({\bf q}, \nu)} 
G_0 ({\bf q} + {\bf k}, \nu + \omega), 
\end{equation}
where $v_q = 2 \pi e^2 /q$ is the bare Coulomb interaction between
electrons (and ${\bf q}$, ${\bf k}$ appropriate 2D wavevectors),
\begin{equation}
\label{eq:G0}
G_0({\bf k}, \omega) =
{1 - n_F(\xi_{\bf k}) \over \omega - \xi_{\bf k} + i\eta } +
{n_F(\xi_{\bf k}) \over \omega - \xi_{\bf k} - i\eta } 
\end{equation}
is the Green's function for free electrons with the noninteracting
energy dispersion $\xi_{\bf k} = {k^2 \over 2 m} - \mu$, $\mu$ the
chemical potential, and $\epsilon({\bf k}, \omega)$ is the dynamic
dielectric function. We use $\eta$ to denote an infinitesimal positive
number, and $n_F(x)$ the Fermi function. At zero temperature, $n_F(x)
=1$ when $x \le 0$ and $0$ otherwise. Within RPA, we have
\begin{equation}
\label{eq:epsilon}
\epsilon({\bf k}, \omega) = 1 - v_q \Pi({\bf k}, \omega)
\end{equation}
with $\Pi({\bf k}, \omega)$ the noninteracting electronic 2D
polarizability (i.e. the bare bubble in Fig.~\ref{fig:feynman}):
\begin{equation}
\label{eq:bubble}
\Pi({\bf k}, \omega) = 2 \int {d^2 q \over (2 \pi)^2} 
{n_F(\xi_{\bf q}) - n_F(\xi_{{\bf q} + {\bf k}}) \over 
\xi_{\bf q} - \xi_{{\bf q} + {\bf k}} + \omega},
\end{equation}
Previous calculation shows~\cite{ando} that in 2D we have
\begin{eqnarray}
\label{eq:2DP}
\Pi ({\bf k}, \omega) = -{m \over \pi} + {m^2 \over \pi k^2} 
\left[ \sqrt{ \left(\omega + {k^2 \over 2m} \right)^2 
- {2 \mu k^2 \over 2m} } \right. \nonumber \\ 
- \left. \sqrt{ \left(\omega - {k^2 \over 2m} \right)^2 
- {2 \mu k^2 \over 2m} }~\right].
\end{eqnarray}
Note that in Eq.~(\ref{eq:bubble}) and Eq.~(\ref{eq:2DP}), the
frequency $\omega$ can be any complex number, and the branch cuts of
the square roots are chosen such that their imaginary part is
positive.

\begin{figure}[htbp]
\centering \includegraphics[width=2in]{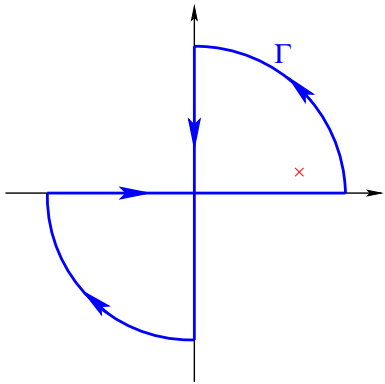}
  \caption{(Color online) Contour in the frequency $\nu$ plane. 
    The cross denotes the possible position of the pole of Green's
    function.}
\label{fig:contour}
\end{figure}

Due to the difficulty with the principal value integration and
singularities of $1 / \epsilon({\bf k}, \omega)$ along the real axis
in Eq.~(\ref{eq:E0}), it is advantageous to follow the standard
procedure and detour the frequency integration from the real axis to
imaginary axis~\cite{quinn}. After choosing the contour of frequency
integration as in Fig.~\ref{fig:contour}, we consider the integration
\begin{equation}
\label{eq:contour}
\oint_{\Gamma} {d \nu \over 2 \pi i} w({\bf q}, \nu) 
G_0 ({\bf q} + {\bf k}, \nu + \omega),
\end{equation}
where $w({\bf q}, \nu)$ is any complex function that is analytic in
the upper and lower half of the complex plane. On the one hand,
(\ref{eq:contour}) equals the integration of the integrand along real
and imaginary axes since the integration along the curve part of the
contour $\Gamma$ (Fig.~\ref{fig:contour}) cancels out. On the other
hand, (\ref{eq:contour}) equals the residue due to the pole of Green's
function
\begin{eqnarray}
\label{eq:residue}
&&( 1 - n_F(\xi_{{\bf q} + {\bf k}}) ) 
\theta(\omega - \xi_{{\bf q} + {\bf k}} )  
w({\bf q}, \xi_{{\bf q} + {\bf k}} - \omega - i \eta) \nonumber \\
&+& n_F(\xi_{{\bf q} + {\bf k}}) 
\theta(\xi_{{\bf q} + {\bf k}} - \omega) 
w({\bf q}, \xi_{{\bf q} + {\bf k}} - \omega + i \eta),
\end{eqnarray}
where $\theta(x) = 1$ when $x > 0$ and $0$ otherwise. Now by setting
\begin{equation}
\label{eq:w}
w({\bf k}, \omega) = {1 \over \epsilon({\bf k}, \omega)} -1,
\end{equation}
we have
\begin{eqnarray}
\label{eq:E1}
\Sigma({\bf k}, \omega) &=& 
- \int {d^2 q \over (2 \pi)^2} v_q n_F(\xi_{{\bf q} + {\bf k}}) )
\nonumber \\
&&- \int {d^2 q \over (2 \pi)^2} 
\int {d \nu \over 2 \pi i} v_q w({\bf q}, \nu) 
G_0 ({\bf q} + {\bf k}, \nu + \omega) 
\nonumber \\
&=& \Sigma^{\mbox{ex}} + \Sigma^{\mbox{res}} + \Sigma^{\mbox{line}},
\end{eqnarray}
where
\begin{eqnarray}
\label{eq:Eex}
&&\!\!\!\!\!\!\!\!\!\!\!\!\!\!\!
\Sigma^{\mbox{ex}}({\bf k}, \omega) = 
- \int {d^2 q \over (2 \pi)^2} v_q n_F(\xi_{{\bf q} + {\bf k}}), \\
&&\!\!\!\!\!\!\!\!\!\!\!\!\!\!\!
\Sigma^{\mbox{res}}({\bf k}, \omega) = 
- \int {d^2 q \over (2 \pi)^2} 
\Big[ \nonumber \\
&&( 1 - n_F(\xi_{{\bf q} + {\bf k}}) ) 
\theta(\omega - \xi_{{\bf q} + {\bf k}} ) 
w({\bf q}, \xi_{{\bf q} + {\bf k}} - \omega - i \eta)  
\nonumber \\ 
&& + n_F(\xi_{{\bf q} + {\bf k}}) 
\theta(\xi_{{\bf q} + {\bf k}} - \omega) 
w({\bf q}, \xi_{{\bf q} + {\bf k}} - \omega + i \eta) 
\Big], \\
&&\!\!\!\!\!\!\!\!\!\!\!\!\!\!\!
\Sigma^{\mbox{line}}({\bf k}, \omega) = 
 - \int {d^2 q \over (2 \pi)^2} \int {d \nu \over 2 \pi}
{w({\bf q}, i \nu) \over i \nu + \omega - \xi_{{\bf q} + {\bf k}}}.
\end{eqnarray}
Note that the static exchange part $\Sigma^{\mbox{ex}}$ is separated
from the dynamical self-energy to improve the numerical convergence of
the line part $\Sigma^{\mbox{line}}$. The exchange (or Fock)
self-energy is obviously frequency independent (but depends strongly
on the wave-vector). There is no Hatree-contribution to the
self-energy in a uniform electron liquid.

It is easy to see that the real part of the self energy $\Sigma'$,
which we need in order to calculate the renormalized effective mass,
can be written as
\begin{eqnarray}
\label{eq:ReE}
&&\!\!\!\!\!\!\!\!\!
\Sigma'({\bf k}, \omega) = 
- \int {d^2 q \over (2 \pi)^2} v_q n_F(\xi_{{\bf q} + {\bf k}}) 
\nonumber \\
&-& \int {d^2 q \over (2 \pi)^2} v_q 
w'({\bf q}, \xi_{{\bf q} + {\bf k}} - \omega) 
[n_B(\xi_{{\bf q} + {\bf k}} - \omega) + n_F(\xi_{{\bf q} + {\bf k}})]
\nonumber \\
&-& \int {d^2 q \over (2 \pi)^2} \int {d \nu \over 2 \pi}
w({\bf q}, i \nu) 
{\omega - \xi_{{\bf q} + {\bf k}} \over 
\nu^2 + (\omega - \xi_{{\bf q} + {\bf k}} )^2},
\end{eqnarray}
where $n_B(x)$ is the Bose function at zero temperature, (i.e.,
$n_B(x) = -1$ when $x < 0$ and $0$ otherwise) $ w'({\bf q}, \xi_{{\bf
    q} + {\bf k}} - \omega) = \mbox{Re}~ w({\bf q}, \xi_{{\bf q} +
  {\bf k}} - \omega)$, and $w({\bf q}, i \nu)$ is real since $w^*({\bf
  q}, i \nu) = w({\bf q}, -i \nu) = w({\bf q}, i \nu)$.  For
completeness, we also mention that the imaginary part of the
self-energy comes only from the residue part of the self-energy
$\Sigma^{\mbox{res}}({\bf k}, \omega)$. The imaginary part of the
self-energy does not play any role in the quasiparticle effective mass
calculation since this is purely a Fermi surface property.


\subsection{Finite width correction}
\label{sec:form:W}

The real 2D systems have a finite layer width in the transverse
direction which reduces the strength of the 2D Coulomb interaction. In
order to test whether our result is model dependent, we include finite
width correction in our calculation to see how the result changes
compared to our ideal 2D results. The origin of this correction comes
from the fact that any 2DES confined within a quantum well or
heterostructure is physically a three dimensional system, with the
width in the confined dimension small but finite. Due to the finite
width of the quasi-2D electron gas system (such as Si MOSFETs and
GaAs/AlGaAs heterojunctions), the Coulomb interaction should not be
written as the ideal 2D form $v_q = 2 \pi e^2 /q$, but should have a
finite-width quantum form factor $F$ such that~\cite{ando}
\begin{equation}
\label{eq:vW}
v_q = {2 \pi e^2 \over \bar{\kappa} q} F(qb),
\end{equation}
where
\begin{equation}
\label{eq:Ffactor}
F(x) = (1 + {\kappa_{\rm ins} \over \kappa_{\rm sc}}) 
{8 + 9x + 3x^2 \over 8 (1+x)^3 }
+ (1 - {\kappa_{\rm ins} \over \kappa_{\rm sc}}) {1 \over 2 (1 + x)^6}.
\end{equation}
with
\begin{equation}
\label{eq:b}
b = \left( {\kappa_{\rm sc} \hbar^2 \over 48 \pi m_z e^2 n^*} \right)^{1/3}
\end{equation}
denoting the width of the quasi 2D electron gas, $\kappa_{\rm sc}$ and
$\kappa_{\rm ins}$ are the dielectric constants for the space charge
layer and the insulator layer, $\bar{\kappa} = (\kappa_{\rm
  sc}+\kappa_{\rm ins})/2$, $m_z$ is the band mass in the direction
perpendicular to the quasi 2D layer, and $n^* = n_{\rm depl} + {11
  \over 32}n$ with $n_{\rm depl}$ the depletion layer charge density
and $n$ the 2D electron density. We choose $n_{\rm depl}$ to be zero
in our calculations. When including finite width correction in our
calculation, we replace $v_q$ by Eq.~(\ref{eq:vW}) in both
Eq.~(\ref{eq:E0}) and Eq.~(\ref{eq:epsilon}). Note that we have chosen
the finite width correction for a triangular potential confinement as
in GaAs heterostructures or Si MOSFETs. In Si MOSFETs system,
\begin{eqnarray}
\label{eq:cSi}
\kappa_{\rm sc} &=& 11.5 ~({\rm Si}),~~~ 
\kappa_{\rm ins} = 3.9 ~({\rm SiO_2}),
\nonumber \\
m &=& 0.19 m_e,~~~~~~ 
m_z = 0.916 m_e,
\end{eqnarray}
while in GaAs heterostructures,
\begin{eqnarray}
\label{eq:cGa}
\kappa_{\rm sc} &=& 12.9 ~({\rm GaAs}),~~~ 
\kappa_{\rm ins} = 10.9 ~({\rm AlGaAs}),~ \nonumber \\
m &=& 0.07 m_e,~~~~~~~~~~~ 
m_z = 0.07 m_e.
\end{eqnarray}


\subsection{Vertex correction}
\label{sec:form:V}

Even though there has been little systematic work going beyond RPA, it
is still possible to include in some approximate way the effects of
vertex corrections to $\Sigma({\bf k}, \omega)$. The purpose here is
again to test whether our numerical result is highly model dependent.
In some of our calculations we therefore approximate vertex
corrections to the self-energy by including local-field corrections of
the Hubbard type in the self-energy and in the dielectric
function~\cite{marmorkos, jonson, rice, book}. In this approximation
the self-energy $\Sigma({\bf k}, \omega)$ in Eq.~(\ref{eq:E0}) is
modified by a vertex term $\gamma({\bf k}, \omega)$:
\begin{eqnarray}
\label{eq:EV}
\Sigma({\bf k}, \omega) &=& - \int {d^2 q \over (2 \pi)^2} 
\int {d \nu \over 2 \pi i} {v_q \gamma(q, \omega) 
\over \epsilon({\bf q}, \nu)} \nonumber \\
&&~~~~~~~\cdot G_0 ({\bf q} + {\bf k}, \nu + \omega), 
\end{eqnarray}
with the dielectric function now also including the same vertex (local
field) correction:
\begin{equation}
\label{eq:epsilonV}
\epsilon({\bf k}, \omega) 
= 1 - v_q \Pi({\bf k}, \omega) \gamma(k, \omega). 
\end{equation}
It is important to emphasize that the vertex correction must appear
both in the self-energy and in the dielectric function for the sake of
consistency. Within RPA, $\gamma \equiv 1$ by definition, and in
general $\gamma({\bf k}, \omega)$ is unknown. We choose
\begin{equation}
\label{eq:gamma}
\gamma(k, \omega) = {1 \over 1 + G(k) v_k \Pi(k, \omega)},
\end{equation}
where $G(k)$ is a static local-field correction given in the 2D
Hubbard approximation by~\cite{jonson}
\begin{equation}
\label{eq:Gfactor}
G(k) = {k \over 2 \sqrt{k^2 + k_F^2}}.
\end{equation}
If the inclusion of this local-field vertex correction substantially
modifies our RPA results, then our qualitative conclusions about the
effective mass divergence become highly suspect. We will see this not
to be the case. Note that in section~\ref{sec:RPA:H} we introduce
another form of the local field factor $G(k)$ (and refer to this
approximation as ``HA2'', in contrast to the $G(k)$ factor given in
Eq.~(\ref{eq:Gfactor}) which we call ``HA1'' approximation) in order
to compare with our RPA results and discuss the qualitative validity
of our effective mass results.


\subsection{Effective mass in on-shell and off-shell approximations}
\label{sec:form:m}

Even after an accurate calculation of $\Sigma({\bf k}, \omega)$, there
are subtle issues associated with the calculation of the quasiparticle
effective mass. The single particle energies are given by the
positions of the poles of the interacting Green's functions and are
determined by the Dyson's equation~\cite{book}
\begin{equation}
\label{eq:offshellE}
E({\bf k}) 
= {k^2 \over 2 m} + \Sigma'({\bf k}, E({\bf k}) - \mu),
\end{equation}
where $\Sigma'$ is the real part of the self-energy. The effective
mass of the quasiparticles on Fermi surface can be written as
\begin{equation}
\label{eq:offshellm}
{m^* \over m} = \left. {1 - {\partial \over \partial \omega} 
\Sigma'({\bf k}, \omega ) \over 
1 + {m \over k} {\partial \over \partial k}
\Sigma'({\bf k}, \omega) } \right|_{k = k_F, \omega = 0},
\end{equation}
where $k_F$ denotes the Fermi momentum. Note that $\omega$ is measured
from the chemical potential $\mu$, which is renormalized by the
electron-electron interaction. $\mu$ can be obtained through solving
the following self-consistent equation
\begin{equation}
\label{eq:mu}
\mu = {k_F^2 \over 2 m} + \Sigma'(k_F, \omega = 0; \mu),
\end{equation}
where $\Sigma'({\bf k}, \omega; \mu)$ is given by Eq.~(\ref{eq:ReE}),
with chemical potential $\mu$ an explicit variable with the {\em
  noninteracting} chemical potential being $\mu \equiv E_F \equiv
k_F^2/2m$, the 2D Fermi energy (i.e. $\Sigma' = 0$). Plugging the
result of renormalized chemical potential $\mu$ into
Eq.~(\ref{eq:offshellm}), we can then calculate the effective mass. We
name this way of effective mass calculation the ``off-shell
approximation'', in contrast to the ``on-shell approximation'' which
we describe in the following.

If the self-energy is calculated only in the lowest order, it is not
sensible to solve Eq.~(\ref{eq:offshellE}) but rather take its first
iteration so that different orders are not mixed in the perturbation
theory:
\begin{equation}
\label{eq:onshellE}
E({\bf k}) 
= {k^2 \over 2 m} + \Sigma'({\bf k}, \xi_{\bf k}).
\end{equation}
This approximation has been used and justified in the
literature~\cite{ting, hedin, rice, dubois}, and here we call it the
``on-shell approximation''. The effective mass can then be written as
\begin{equation}
\label{eq:onshellm}
{m^* \over m} = \left. {1 \over 1 + {m \over k} {d \over d k}
\Sigma'({\bf k}, \xi_{\bf k}) } \right|_{k = k_F},
\end{equation}
where we use the un-renormalized (i.e. the noninteracting) chemical
potential $E_F$ in the right hand side of the above equation.

On-shell and off-shell approximations obviously converge as $r_s \to
0$, where the interaction is weak and the higher-order iterations in
Eq.~(\ref{eq:offshellE}) vanish making Eqs.~(\ref{eq:offshellE}) and
(\ref{eq:onshellE}) as well as Eqs.~(\ref{eq:offshellm}) and
(\ref{eq:onshellm}) completely equivalent. But for $r_s > 1$, where
the system is strongly interacting, there is a very large difference
between the two approximations which increases with increasing $r_s$.
The fact that on-shell and off-shell approximations give increasingly
different (in fact, qualitatively different in our case as we will see
later in section~\ref{sec:results}) results with increasing $r_s$
calls for a careful examination of the theoretical issue underlying
these two alternate methods for calculating the quasiparticle
effective mass (see section~\ref{sec:RPA} for more details on this
topic) to decide which one is the correct approximation to use in a
particular situation (and why). One could take the easy way out and
insist that, as often is the case in solid state physics, empiricism
should decide which of these approximations is ``better'' by comparing
with experimental data. (By this empirical criterion of agreement with
experiment, the on-shell approximation certainly wins out as the
decisively better approximation for the 2D effective mass calculation
since the off-shell RPA self-energy approximation for the 2D effective
mass produces essentially a nearly density-independent effective mass
with a rather small $\sim 10 - 20\%$ mass renormalization for $r_s
\gtrsim 10$, which is in sharp contrast with the recent experimental
claims~\cite{kravchenko1, kravchenko2, kravchenko3, pudalov}.)  We
feel, however, that the issue of on-shell versus off-shell
approximation is sufficiently important to merit serious theoretical
considerations on its own without resorting to empiricism. Arguments
supporting the on-shell RPA scheme (over the off-shell one) were
actually provided in rather compelling terms by Dubois~\cite{dubois}
and by Rice~\cite{rice} a long time ago (see also more recent
references~\cite{vinter} and~\cite{ting} for the corresponding 2D
arguments). The basic argument, which we will expand upon and discuss
in more detail in section~\ref{sec:RPA}, is that of mixing orders in a
systematic perturbation expansion. There is no question that if one
calculates the {\em exact} self-energy (an obviously impossible task
in our problem) then one must solve the Dyson's equation to obtain the
interacting quasiparticle Green's function leading automatically to
the off-shell effective mass formula (i.e. Eqs.~(\ref{eq:offshellE})
and (\ref{eq:offshellm})). The problem arises from the leading-order
perturbative (in the dynamically screened interaction) nature of the
RPA self-energy (formally $\Sigma \sim u G_0$ in the RPA theory, i.e.
$\Sigma$ is in the leading order of $u$ and $G_0$), which implies that
all quantities should be expanded only to the leading order in the
interaction; otherwise one is mixing orders in an inconsistent manner
in the theory rendering the theory highly suspect. This means that
Eq.~(\ref{eq:offshellE}) for the quasiparticle energy dispersion
$E({\bf k})$ should {\em only} be solved to the leading order in
$\Sigma$ (i.e. keeping only its first iteration, as in the on-shell
approximation of Eq.~(\ref{eq:onshellE})), and {\em not} to all orders
as in the off shell approximation. This is explicitly seen by
iterating Eq.~(\ref{eq:offshellE}) in an order by order manner leading
to
\begin{eqnarray}
\label{eq:Eiter}
E({\bf k}) = {k^2 \over 2m} + \Sigma'({\bf k}, \xi_k) + \Big[ \Sigma' 
\left({\bf k}, \xi_k + \Sigma'({\bf k}, \xi_k) \right) \nonumber \\
+\Sigma' \big({\bf k}, \xi_k + \Sigma' \left({\bf k}, \xi_k 
+ \Sigma'({\bf k}, \xi_k) \right) \big) + \cdots \Big].
\end{eqnarray}
Clearly all the terms inside the square bracket are higher orders in
the interaction $u$, and should not be kept (as long as $\Sigma'$ is
obtained in the leading order $G_0 u$ RPA approximation) for the sake
of consistency, leaving us with the on-shell approximation of
Eq.~(\ref{eq:onshellE}). It is important to emphasize that the
distinction between off-shell and on-shell approximations disappears
in the weakly interacting $r_s \to 0$ high-density limit since all the
terms inside the square bracket are higher-order in $r_s$ and
disappear as $r_s \to 0$. We discuss the issue of on-shell and
off-shell approximations further in section~\ref{sec:RPA} where the
justification for (or the validity of) our RPA self-energy
approximation (at large $r_s$) is taken up.


\section{Results}
\label{sec:results}

\subsection{Effective mass within on-shell approximation}
\label{sec:results:onshell}

We first present the results of our effective mass calculation using
Eq.~(\ref{eq:onshellm}). We use the dimensionless quantity $r_s$ to
denote the density of the system, $r_s = m e^2 /(\hbar^2 \sqrt{\pi
  n})$, showing our calculated results as a function of $r_s$ where
$n$ is the electron density. We concentrate on the high $r_s$ (i.e.
low electron density) behavior of the effective mass. We do mention,
however, that we have calculated $m^*(r_s)$ for all values of $r_s$
reproducing the earlier results~\cite{vinter, ting, jalabert,
  marmorkos, giuliani1, giuliani2} in the literature in the $r_s < 5$
regime. In particular, we quantitatively reproduce the analytical 2D
behavior ${m^* \over m} = 1 - {r_s \over 2 \sqrt{2}} \ln {\sqrt{2}
  \over r_s}$ in the $r_s \to 0$ limit within our numerical
calculations.

\begin{figure}[htbp]
\centering \includegraphics[width=3in]{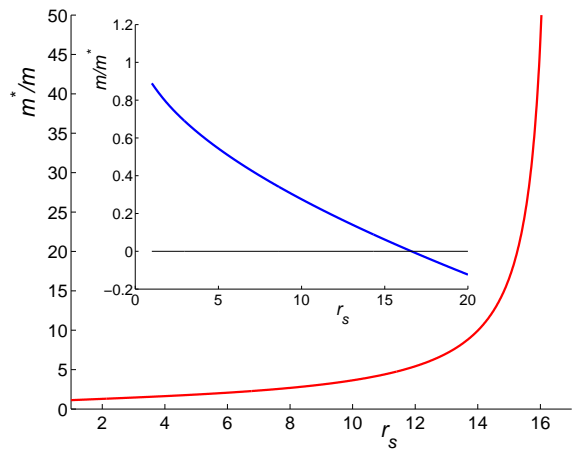}
  \caption{(Color online) Calculated 2D RPA effective mass within 
    on-shell approximation as a function of $r_s$. Inset: the
    denominator of expression~(\ref{eq:onshellm}) at different $r_s$,
    which equals $m/m^*$. It is clear that the effective mass diverges
    at $r_s \sim 16.6$ within the on-shell RPA approximation for the
    ideal 2D system.}
\label{fig:onshell}
\end{figure}

From Fig.~\ref{fig:onshell} we can see that the 2D effective mass has
a divergence at low carrier densities ($r_s \sim 16.6$). Examining
Eq.~(\ref{eq:onshellm}) carefully, it is not hard to identify the
origin of this divergence. The dimensionless term ${d \over d k}
\Sigma'({\bf k}, \xi_{\bf k})|_{k = k_F}$ in the denominator of
Eq.~(\ref{eq:onshellm}) is a negative quantity for large $r_s$, and
our numerical calculation shows that its magnitude increases as $r_s$
increases. At $r_s \sim 16.6$, the magnitude of this term passes
through unity, which means that the calculated effective mass
diverges.

\begin{figure}[htbp]
\centering \includegraphics[width=3in]{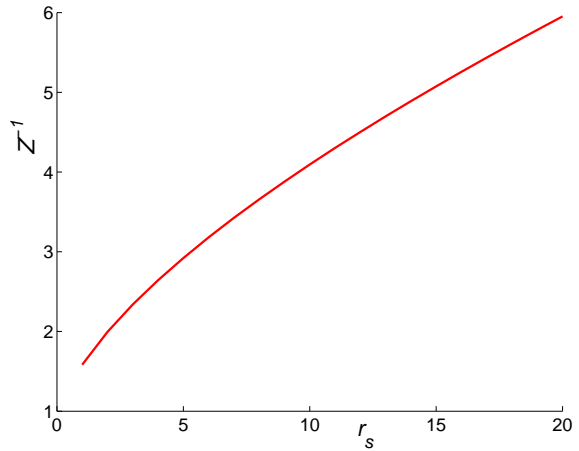}
  \caption{Calculated inverse $Z$ factor as a function of $r_s$ 
    values. $Z^{-1} \equiv \left[1 - \partial_\omega \Sigma({\bf k},
    \omega) \right]_{k = k_F, \omega = 0}$}
\label{fig:zfactor}
\end{figure}

We point out that the effective mass divergence we find does {\em not}
in any way automatically imply a concomitant failure of the Fermi
liquid theory (of course, we do {\em not} know what happens at $r_s$
values above the critical density for the mass divergence). In
particular, in Fig.~\ref{fig:zfactor} we present our calculated
inverse Fermi liquid renormalization factor $Z^{-1} \equiv \left[1 -
\partial_\omega \Sigma({\bf k}, \omega) \right]_{k = k_F, \omega = 0}$
as a function of $r_s$. We see that $Z$ remains well-behaved around
where $m^*/m$ diverges (i.e. $Z^{-1}$ does not manifest any
divergence). Since $Z$ denotes the Fermi surface discontinuity at
$k_F$, $Z \ne 0$ means that the Fermi liquid behavior is not
destroyed.

\begin{figure}[htbp]
\centering \includegraphics[width=3in]{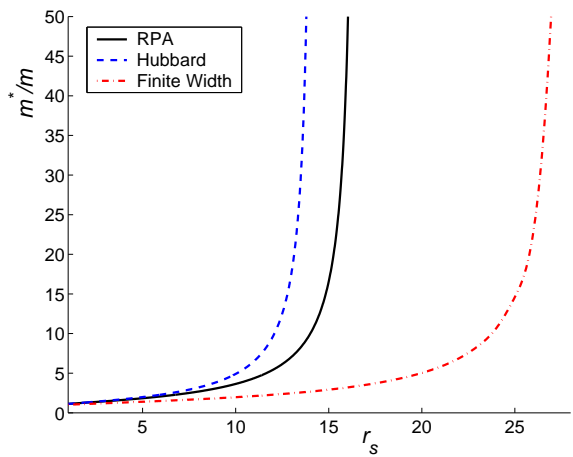}
  \caption{(Color online) Calculated effective mass within the 
    on-shell approximation at different $r_s$ values. The solid
    (black) curve is the pure RPA result, the dashed (blue) curve is
    the ideal 2D result with Hubbard vertex correction (HA1), and the
    dot dashed (red) curve is the RPA result with finite width
    correction. Note that all three curves show effective mass
    divergence. In the ideal 2D Hubbard approximation result, $m^*$
    diverges at $r_s \sim 14.2$, while in the finite width RPA result,
    $m^*$ diverges at $r_s \sim 27.8$.  Note that in the finite width
    correction we have set $\kappa_{\rm ins} = \kappa_{\rm sc} =
    \bar{\kappa}$ and $m_z = m$ (with $m$ being the in-plane band
    mass), and the correction is much stronger than in real Si MOSFET
    quasi-2D system case as shown in Fig.~\ref{fig:real}.}
\label{fig:mrCorr}
\end{figure}

To test whether our result is highly model dependent, we introduce two
kinds of corrections as we explained in section~\ref{sec:form:W} and
section~\ref{sec:form:V}. Fig.~\ref{fig:mrCorr} shows the calculated
effective mass including finite width corrections with RPA, and the
Hubbard vertex correction in the ideal 2D case. We find that finite
width correction suppresses the effective mass renormalization, while
vertex correction enhances it at large $r_s$ values. The important
point is that both of these corrections show clear effective mass
divergence -- for the finite width correction the critical $r_s$ for
the divergence goes up since the bare interaction is softened by
finite width effect whereas vertex correction actually decreases the
critical $r_s$ since local field correction tends to enhance
interaction effects. In fact the inverse mass passes zero as $r_s$
goes beyond critical values in each of these two cases, and the
qualitative behavior of $m^*(r_s)$ is very similar to that shown in
Fig.~\ref{fig:onshell}. Note that in our effective mass calculation
with finite width correction (results shown in Fig.~\ref{fig:mrCorr}),
we have set $\bar{\kappa} = \kappa_{\rm sc} = \kappa_{\rm ins}$ and
$m_z = m$, which is approximately applicable for GaAs hetrostructure
systems. We also calculate within the quasi-2D electron gas picture
the effective mass of Si MOSFETs as a function of carrier density.
(These results are shown in Fig.~\ref{fig:real} and discussed in
section~\ref{sec:con}.)

\begin{figure}[htbp]
\centering \includegraphics[width=3in]{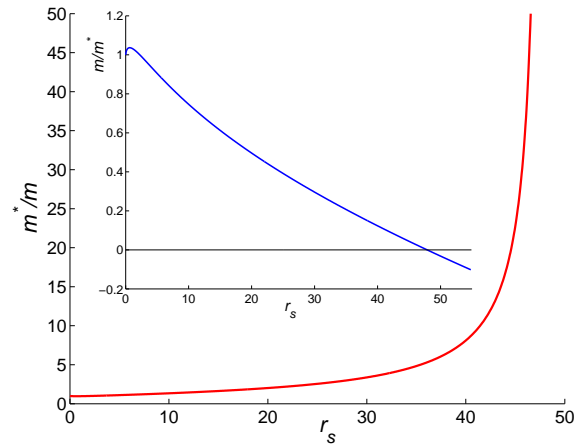}
  \caption{(Color online) Calculated 3D RPA effective mass within 
    on-shell approximation as a function of $r_s$. Inset: the
    denominator of expression~(\ref{eq:onshellm}) at different $r_s$,
    which equals $m/m^*$. It is clear that the effective mass diverges
    as $r_s \sim 48$ within the on-shell RPA approximation for the 3D
    system. }
\label{fig:onshell3D}
\end{figure}

It is reasonable to ask whether the low density effective mass
divergence shown in Fig.~\ref{fig:onshell} is a specific
characteristic of strongly interacting 2D electron systems or is a
generic property of interacting electron systems. To answer this
important question we have carried out a similar calculation for a 3D
electron system interacting via 3D Coulomb interaction $v_q = 4 \pi
e^2 /q^2$, finding a very similar effective mass divergence at large
$r_s$ (see Fig.~\ref{fig:onshell3D}). The divergence of $m^*(r_s)$ at
some large critical $r_s$ is thus a generic feature (similar to He-3)
of strongly interacting electron systems with system dimensionality
(2D or 3D) playing no role.

\subsection{Self consistently calculated chemical potential} 
\label{sec:results:mu}

The off-shell approximation effective mass calculation requires the
chemical potential $\mu$ to be calculated self-consistently using
Eq.~(\ref{eq:mu}).

\begin{figure}[htbp]
\centering \includegraphics[width=3in]{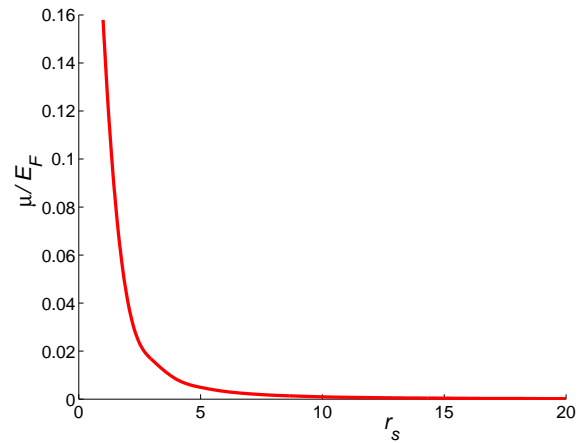}
  \caption{Calculated renormalized chemical potential $\mu$ using 
    Eq.~(\ref{eq:mu}). $E_F$ denotes the Fermi energy for the
    corresponding non-interacting Fermi system.}
\label{fig:mu}
\end{figure}

In Fig.~\ref{fig:mu} we present our numerical result for the
renormalized chemical potential. The result shows that the chemical
potential decreases and approaches $0$ as $r_s$ gets larger. We can
also see that ${d \mu \over d r_s}$, or the inverse compressibility
$\kappa^{-1} = n^2 {d \mu \over d n}$ approaches zero as density
decreases or $r_s$ increases.


\subsection{Effective mass within the off-shell approximation}
\label{sec:results:offshell}

\begin{figure}[htbp]
\centering \includegraphics[width=3in]{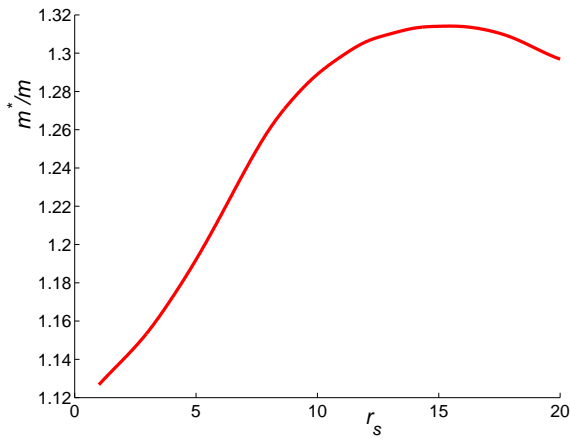}
  \caption{Calculated RPA effective mass within off-shell 
    approximation using Eq.~(\ref{eq:offshellm}).}
\label{fig:offshell}
\end{figure}

Using the renormalized chemical potential, we obtain the off-shell
effective mass results, which are presented in
Fig.~\ref{fig:offshell}. Our off-shell results differ qualitatively
from our on-shell results, and no mass divergence occurs within the
off-shell RPA calculation at any $r_s$ values. In fact the effective
mass decreases at high $r_s$, which is in qualitative disagreement
with the experimental results~\cite{smith, coleridge, pan,
  kravchenko1, kravchenko2, pudalov}.  It is also strange that the
calculated $m^*(r_s)$ in Fig.~\ref{fig:offshell} (i.e., the off-shell
RPA) shows extremely weak many-body renormalization, typically less
than $30\%$ even for $r_s$ as large as $10 - 20$.

We now provide some rather simple (essentially hand-waving) arguments
to better understand the qualitative difference between the results of
on- and off-shell approximations. We examine the $r_s$ dependence of
the partial derivatives of the self-energy $\partial_\omega
\Sigma'({\bf k}, \omega)|_{(k = k_F,~\omega = 0)}$ and $(m/k)
\partial_k \Sigma'({\bf k}, \omega)|_{(k = k_F,~\omega = 0)}$
appearing in Eq.~(\ref{eq:offshellm}). Both of these quantities
increase in their magnitude with $r_s$. As an oversimplified model, we
make the approximation that $ \partial_\omega \Sigma'({\bf k},
\omega)|_{(k = k_F,~\omega = 0)} = - \alpha r_s$ and $(m/k) \partial_k
\Sigma'({\bf k}, \omega)|_{(k = k_F,~\omega = 0)} = \beta r_s$, where
$\alpha \gtrapprox \beta > 0$ since our numerical results show that
the two partial derivatives are opposite in sign and close in
magnitude. Off-shell calculation for effective mass using
Eq.~(\ref{eq:offshellm}) yields the result $(1 + \alpha r_s) / (1 +
\beta r_s)$, which increases with $r_s$ at first saturating at large
$r_s$ without any divergence. On the other hand, the on-shell
approximation has both of these two derivatives appearing in the
denominator of the effective mass expression (see
Eq.~(\ref{eq:onshellm}), which yields $1 /[1 - (\alpha - \beta) r_s]$.
(Note that in this oversimplified argument we ignore the difference
between the renormalized chemical potential and non-interacting
chemical potential.) Now we can manifestly see that on- and off- shell
RPA approximation give the same $m^*(r_s)$ in the $r_s \to 0$ limit.
However, as $r_s$ increases the on-shell effective mass keeps on
increasing, and diverges at $r_s \sim 1/(\alpha - \beta)$, whereas the
off-shell effective mass essentially saturates for $r_s > \alpha (>
\beta)$, with a rather small ($= \alpha / \beta< 30 \%$)
renormalization. We have already argued that within the leading order
``$G_0 u$'' RPA self-energy approximation, the on-shell effective mass
calculation is more consistent (and hence, better) than the off-shell
approximation. We will discuss this issue later in
section~\ref{sec:RPA} again.

\begin{figure}[htbp]
\centering \includegraphics[width=3in]{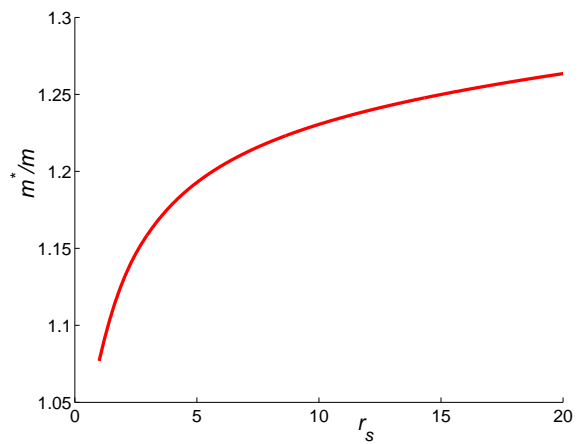}
  \caption{Calculated 2D RPA effective mass using
    Eq.~(\ref{eq:offshellm}), while keeping $\mu = E_F$.}
\label{fig:mixed}
\end{figure}

For completeness, we also mention that often in the literature the
effective mass is calculated using Eq.~(\ref{eq:offshellm}) and the
noninteracting chemical potential rather than interacting $\mu$ (i.e.
$\mu = E_F$ is used in Eq.~(\ref{eq:offshellm}) ). This is somewhat a
mixture of our on-shell and off-shell approximations as it uses
noninteracting $\mu$ but self-consistent quasiparticle energy, and is
logically inconsistent. We have also calculated effective mass using
this method, and the results shown in Fig.~\ref{fig:mixed} are similar
to the off-shell results of Fig.~\ref{fig:offshell}.

\section{Validity of RPA}
\label{sec:RPA}

We have discussed somewhat the validity of the RPA self-energy
approximation in the context of our quasiparticle effective mass
calculation. In this section we concentrate entirely on this important
theoretical issue, and discuss it at some depth providing several
complementary arguments supporting (at least) the qualitative validity
of RPA even at large $r_s$ value of primary interest to us in this
work.

We start with some general observations about RPA. In the context of
the quasiparticle self-energy (and effective mass) calculations it
implies three distinct steps of approximations: (i) Calculate the
self-energy in a single loop ``$G_0 u$'' approximation (often called
``GW'' approximation in the literature following Hedin's influential
publication~\cite{hedin} dating back forty years although the first
such self-energy calculation was originally carried out by Quinn and
Ferrell~\cite{quinn}), i.e. in the leading order dynamically screened
interaction; (ii) Obtain the dynamically screened interaction by using
the infinite series of bare bubble diagrams, i.e. approximate the
irreducible polarizability by the noninteracting polarizability ``$G_0
G_0$'' neglecting all vertex and self-energy corrections to the
polarizability; (iii) Calculate the effective mass using the on-shell
approximation based on the rationale that in a leading order
calculation of the self-energy the quasiparticle energy should only be
calculated in the leading-order approximation to the Dyson's
(integral) equation so as not to mix various perturbative orders
arbitrarily and inconsistently. Of these three steps in the RPA
effective mass calculation, the first two are essential approximations
because going beyond the leading order approximation in the
dynamically screened interaction and beyond the bare noninteracting
polarizability (``the bare bubble'') in dynamical screening are
essentially impossible in calculating the full frequency and
wave-vector dependent self-energy $\Sigma({\bf k}, \omega)$ necessary
for obtaining $m^*(r_s)$. The third approximation, the on-shell
approximation defined in step (iii) above, is, as explained in
section~\ref{sec:results}, a nonessential approximation which we make
because we believe it to be a better approximation than the full
off-shell solution of Dyson's equation.

We first assert that it is incorrect to think of RPA as an expansion
in the $r_s$ parameter as is commonly done. RPA is an expansion in the
dynamically screened Coulomb interaction (i.e. the infinite series of
ring or bubble diagrams as in Fig.~\ref{fig:feynman}), {\em not} a
density expansion in any sense. For example, the RPA calculation for
the plasmon dispersion is exact in the $q \to 0$ limit for {\em all}
$r_s$, and in reality, the RPA-calculated 2D plasmon dispersion
provides an accurate quantitative description for the experimental 2D
plasmon dispersion ~\cite{hwang, pinczuk1, pinczuk2} up to very low
densities ($r_s > 10$). (Note that the RPA plasmon dispersion is
defined entirely by the polarizability through the equation
$\epsilon({\bf q}, \omega) = 0$, i.e. $1 - v_q \Pi({\bf q}, \omega) =
0$. ) What is true is that the RPA self-energy approximation (i.e.
Fig.~\ref{fig:feynman}) becomes {\em exact} in the $r_s \to 0$ limit,
where the quasiparticle effective mass can be written as an expansion
in $r_s$, but the precise expansion parameter even in this $r_s \to 0$
limit is {\em not} $r_s$, but $\alpha = e^2/(\hbar v_F)$ where $v_F$
is the Fermi velocity. Writing the expansion parameter $\alpha$ out in
terms of $r_s$, we see that it is incorrect to think of $r_s$ as the
relevant expansion parameter for the quasiparticle effective mass
calculation even in the well-studied $r_s \to 0$ limit since $\alpha
\approx r_s/12$ (3D), $r_s/4.5$ (2D); and therefore the true expansion
parameter is much less than $r_s$, making the radius of convergence of
an $r_s$-expansion much larger than unity (perhaps qualitatively
explaining why RPA works so well for 3D metals where $r_s \approx 3 -
6$). What is undoubtedly true is that for arbitrary $r_s (>1)$, RPA is
{\em not} exact by any means as it is in the unphysical $r_s \to 0$
limit. This statement is, however, not equivalent to claiming that RPA
is necessarily a very bad approximation for large $r_s$. The argument
for the failure of RPA at large $r_s$ is based entirely on the wrong
premise that RPA is a systematic $r_s$-expansion even at large $r_s$.

It has long been known, since the seminal work of Gellmann and
Brueckner fifty years ago~\cite{gellmann}, that RPA (i.e. the ring or
bubble diagrams) provides the asymptotically exact ground state energy
for an interacting electron gas in the high-density $r_s \to 0$
limit~\cite{book}. This is true both in 2D and 3D interacting electron
systems (but {\em not} in 1D system where interaction effects are
nonperturbabtively singular and one must use entirely different
approaches) and leads to a ground state energy expression of the form
$\epsilon_{\rm corr} = A^{(3)} + B^{(3)} \ln r_s~(A^{(2)} + B^{(2)}
r_s \ln r_s$) in 3D (2D) electron liquids in the $r_s \to 0$ limit
(where A and B are known numerical constants), where $\epsilon_{\rm
  corr}$ is only the non-trivial interaction or the correlation part
of the ground state energy per particle (i.e. leaving out the
noninteracting kinetic energy and the Fock exchange energy which goes
as $r_s^{-2}$ and $r_s^{-1}$ respectively in both 2D and 3D systems).
It is important to emphasize that the interaction energy is manifestly
non-analytic around $r_s \to 0$ (due to the $\ln r_s$ term) and
therefore RPA is {\em not} a simple power series expansion in
inverse-density even in the high density $r_s \to 0$ limit. We
emphasize here a surprising (and apparently little-known)
fact~\cite{iwata}: RPA (i.e. the ring diagram approximation) again
gives an asymptotically excellent result for the ground state
correlation energy in the extreme low-density $r_s \to \infty$ limit
as well. Although this was theoretically demonstrated~\cite{iwata} by
Iwata a long time ago, this fact seems to be not very well-known in
the literature.

We show in \ref{sec:RPA:para}, using a very crude and rather
simplistic power counting arguments, that RPA is most likely {\em not}
an expansion in the $r_s$-parameter for arbitrary $r_s$ values, but it
becomes equivalent to an $r_s$ expansion as $r_s \to 0$. In
section~\ref{sec:RPA:V} we revisit the on-shell versus the off-shell
approximation question for the quasiparticle effective mass
calculation on a more formal level, justifying on a diagrammatic basis
the superiority of the on-shell approximation over the off-shell one
showing that the on-shell approximation in fact systematically (albeit
implicitly) includes some higher-order diagrams in the effective
interaction making it therefore better than just the simple
single-loop ``$G_0 u$'' approximation.  Finally, in
section~\ref{sec:RPA:H} we explicitly calculate some appropriate
vertex corrections to the 2D self-energy by including different simple
local field corrections to the bare polarizability showing that at
least this level of an approximate inclusion of vertex corrections to
the theory does not qualitatively change the 2D effective mass results
compared with the RPA effective mass calculations, providing some
additional justification for using the RPA self-energy at large $r_s$
value. None of our arguments establishes the RPA effective mass
calculation as a theoretically rigorous approximation for large $r_s$
(and surely it is not), but we believe that our arguments are
persuasive in demonstrating that this is not necessarily a
qualitatively wrong approximation at large $r_s$.


\subsection{The effective expansion parameter $\delta$ in RPA}
\label{sec:RPA:para}
To approximately figure out the effective expansion parameter in RPA
we reconsider the Feynman-Dyson diagrammatic expansion for the
electron self-energy and the irreducible polarizability in
Fig.~\ref{fig:feynman2} by concentrating on the higher-order diagrams
left out of RPA. From Fig.~\ref{fig:feynman2} we see that in each
higher order (remembering that the expansion is in the screened
interaction, not the bare Coulomb interaction which diverges in the $q
\to 0$ limit) we get an additional factor of $\delta \equiv \left[ G_0
G_0 u {d^2 q d \omega \over (2 \pi)^3} \right]$ with the formal series
for the polarizability and the self-energy being:
\begin{eqnarray}
\label{eq:series}
\Pi &\Rightarrow& 
G_0 G_0 + G_0 (G_0 G_0 u) G_0 + G_0 (G_0 G_0 u G_0 G_0 u) G_0 
\nonumber \\
&&+ \cdots, \nonumber \\
\Sigma &\Rightarrow& 
G_0 u + G_0 u (G_0 G_0 u) + G_0 u (G_0 G_0 u G_0 G_0 u) 
\nonumber \\
&&+ \cdots, 
\end{eqnarray}
where the first terms $G_0 G_0$ (for $\Pi$) and $G_0 u$ (for $\Sigma$)
are what we keep in RPA (see Fig.~\ref{fig:feynman}).

We now carry out a simple heuristic power counting of the factor $G_0
G_0 u {d^2 q d \omega \over (2 \pi)^3}$ to get:
\begin{eqnarray}
\label{eq:pwrcnt}
\delta &\equiv& \left[ G_0 G_0 u {d^2 q d \omega \over (2 \pi)^3} 
\right] \nonumber \\ &\Rightarrow&
{1 \over E_F} {1 \over E_F} \left( {2 \pi e^2 \over k_F + q_{TF}}
\right) {(\pi k_F^2) E_F \over (2 \pi)^3},
\end{eqnarray}
where we have used simple dimensional forms for Green's function $G_0
\sim E_F^{-1}$, 2D screened interaction $u \sim 2 \pi e^2 (k_F +
q_{TF})^{-1}$; and phase space $d^2 q \sim \pi k_F^2$; $d \omega \sim
E_F$. This leads to
\begin{equation}
\label{eq:pwrcnt2}
\delta \sim {\sqrt{2} r_s \over 4 \pi (1 + \sqrt{2} r_s)}.
\end{equation}
Thus the 2D RPA expansion parameter $\delta$ is essentially $r_s$ as
$r_s \to 0$, but is actually of the form $r_s /[4 \pi (r_s +
1/\sqrt{2})] r_s)$ for arbitrary $r_s$, which is certainly not
equivalent to an $r_s$ expansion except in the high-density $r_s \ll
1$ limit. Thus, this admittedly crude power counting dimensional
argument leads to the conclusion that RPA remains approximately valid
even at large $r_s$, where $\delta \sim 1/(4 \pi)$ is indeed a small
parameter. This power counting argument holds equally well in 3D
systems where
\begin{equation}
\label{eq:pwrcnt3D}
\delta \equiv G_0 G_0 u \sim 
{r_s (1/6 \pi) (4 / \pi) (4 / 9 \pi)^{1/3} \over
1 + r_s (4 / \pi) (4 / 9 \pi)^{1/3}}
\end{equation}
using the 3D screened Coulomb interaction and 3D phase space integral.
Thus, in 3D, $\delta \sim r_s$ in the $r_s \to 0$ limit, but for $r_s
\gg 1$, $\delta \sim 1/(6 \pi)$ which is small. (We note as an aside
that our crude arguments fail completely if there is a quantum phase
transition to a new ground state at some finite values of $r_s$ since
then the form for $G_0$, $u$ etc. will change qualitatively.)

The crude dimensional arguments of this section, like all power
counting arguments, are too simplistic to be taken rigorously. But,
these dimensional arguments compellingly demonstrate that, contrary to
the very widespread popular belief, RPA is an expansion in the
$r_s$-parameter {\em only} in the $r_s \to 0$ limit but not for
arbitrary values of $r_s$. In fact, for $r_s > 1$, RPA may very well
be reasonably well-valid (as in commonly found on empirical grounds)
because the expansion parameter $\delta$ is not necessarily large just
because $r_s \gg 1$, but is in fact quite small ($\sim 0.1$)
numerically. We emphasize that the dimensional argument given here is
for too simplistic to be taken rigorously -- our purpose here is to
demonstrate that at large $r_s$ the RPA is not equivalent to an $r_s$
expansion. In particular, one serious flaw of our dimensional argument
is that it completely fails to produce the $\ln r_s$ terms which are
known to be present in the $r_s \to 0$ limit, but the very presence of
these logarithmic terms in the high-density limit underscore the fact
that RPA is not a simple power series expansion in $r_s$, which is the
main point we are making here.


\subsection{Approximate higher-order corrections through the RPA on-shell
  approximation}
\label{sec:RPA:V}

\begin{figure}[htbp]
\centering \includegraphics[width=3in]{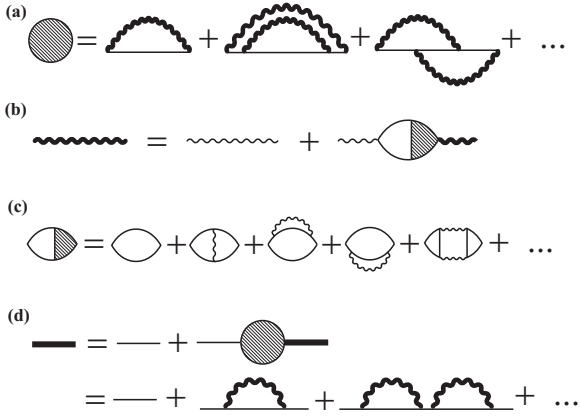}
  \caption{Feynman diagrams for self-energy including higher order 
    terms in the perturbation theory which are excluded in
    Fig.~\ref{fig:feynman}.}
\label{fig:feynman2}
\end{figure}

In this sub-section we show that the on-shell approximation to the
quasiparticle energy, discussed in section~\ref{sec:form}, actually
implicitly incorporates some higher-order corrections (i.e.  beyond
the single-loop $G_0 u$ self-energy of RPA), making it therefore an
improvement over the off-shell approximation involving the full
solution of the Dyson's (integral) equation. The basic ideas
underlying this argument go back a very long time to
Dubois~\cite{dubois} and to Rice~\cite{rice}. We outline the arguments
here in the context of the 2D RPA self-energy calculation since the
arguments seem {\em not} to be very well-known (or well-appreciated)
in the literature. The iterative (off-shell) solution of the Dyson's
equation to the quasiparticle energy can be formally written as
(Fig.~\ref{fig:feynman2}(d)):
\begin{eqnarray}
\label{eq:seriesE}
E_{\bf k} &=& \xi_{\bf k} + \Sigma'({\bf k}, E_{\bf k}) \nonumber \\
&=& \xi_{\bf k} + \Sigma'({\bf k}, \xi_{\bf k}) + \Sigma'({\bf k},
\xi_{\bf k} + \Sigma'({\bf k}, \xi_{\bf k})) + \cdots \nonumber \\
&\approx& \xi_{\bf k} + \Sigma'({\bf k}, \xi_{\bf k}) 
+ \Sigma'({\bf k}, \xi_{\bf k}) 
{\partial \Sigma \over \partial \xi_{\bf k}} + \cdots,
\end{eqnarray}
where we have only explicitly considered up to the second-order in
iteration. One can now formally combine the second order iteration in
the self-energy in the Dyson's equation with the second-order terms in
the diagrammatic perturbative expansion (Fig.~\ref{fig:feynman2}(a))
of $\Sigma$ itself (i.e. the two second order diagrams in
Fig.~\ref{fig:feynman2}(a) beyond the $G_0 u$ single-loop
leading-order RPA self-energy term) to obtain the second order
correction to the basic `on-shell' RPA self-energy term:
\begin{equation}
\label{eq:E2nd}
\Sigma^{(2)}({\bf k}, \omega) \equiv \Sigma^{(2)}_1 + \Sigma^{(2)}_2 
+ {\partial \Sigma^{(1)} \over \partial \xi_{\bf k}} \Sigma^{(1)}
\end{equation}
where $\Sigma^{(2)}_1$ and $\Sigma^{(2)}_2$ are respectively the
contributions from the second order exchange self-energy and the
second order vertex correction diagram, and $\Sigma^{(1)}$ is the
first-order RPA self-energy obtained from the second-order Taylor
expansion of the Dyson's equation expansion.

It is now easy to see that the second order exchange term
($\Sigma^{(2)}_1$) and the term arising from the iteration of the
leading-order self-energy essentially cancel each other:
\begin{eqnarray}
\label{eq:cancel}
\Sigma^{(2)}_1 &+& {\partial \Sigma^{(1)} \over \partial \xi_{\bf k}}
\Sigma^{(1)} = \int {d^2 q d^2 q' \over (2 \pi)^4} 
\int {d \nu d \nu' \over (2 \pi)^2} \nonumber \\
&& \times G_0({\bf k} + {\bf q}, \xi_{\bf k} + \nu)^2 
G_0({\bf k} + {\bf q}', \xi_{\bf k} + \nu') \nonumber \\
&& \times u(q, \nu) 
\left[ u({\bf q} - {\bf q}', \nu' - \nu) - u({\bf q}', \nu') \right].
\end{eqnarray}
By generalizing these arguments order by order, it can be shown in a
straightforward fashion that such approximate cancellations between
the higher-order exchange corrections and the iterations of the
Dyson's equation for the leading-order (i.e. RPA) self-energy occur to
all orders. This leads to the conclusion, already made in
section~\ref{sec:form} and \ref{sec:results}, that the on-shell RPA
self-energy is a better approximation than the off-shell one since the
on-shell approximation to the leading-order RPA self-energy implicitly
contains some aspects of the higher-order corrections through the
order-by-order cancellation demonstrated above. In particular, the
on-shell `$G_0 u$' approximation simulates the self-consistent `$G u$'
approximation , where $G$ is the full interacting Green's function
rather than just the noninteracting $G_0$.

This reinforces our earlier assertion that the RPA on-shell effective
mass calculation is {\em not} a simple high-density expansion, and
could have substantial qualitative (and perhaps even quantitative)
validity at larger $r_s$ values although the extent of its
quantitative validity is difficult to assess because it is {\em not} a
controlled perturbative theory at large $r_s$.

Similar to the other arguments made in this section, we emphasize that
the cancellation between the higher order diagrams
(Fig.~\ref{fig:mrCorr2}) and the Taylor expansion discussed here is
certainly partial and approximate, and again this is by no means a
rigorous argument.


\subsection{Including vertex corrections through Hubbard local field effects}
\label{sec:RPA:H}

\begin{figure}[htbp]
\centering \includegraphics[width=3in]{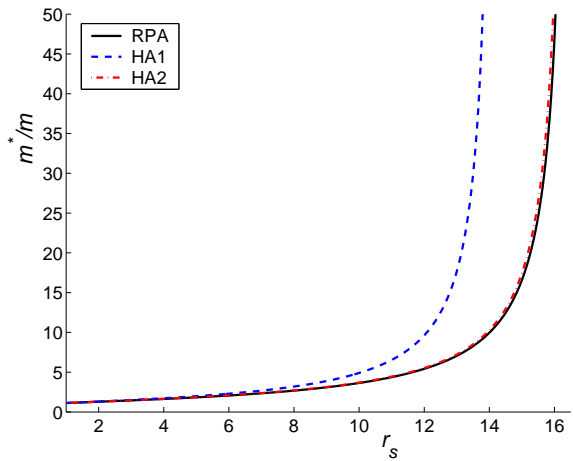}
  \caption{(Color online) Calculated 2D effective mass within the 
    on-shell approximation at different $r_s$ values. The solid
    (black) curve is the pure RPA result, and dashed (blue) curve is
    the HA1 result, and the dot dashed (red) curve is the HA2 result.
    RPA, HA1 and HA2 results all show effective mass divergence. It is
    clear that the RPA result and the HA2 result are strikingly close
    to each other.}
\label{fig:mrCorr2}
\end{figure}

To obtain some ideas about the model dependence of our calculated
quasiparticle effective mass we have already shown results (see
Fig.~\ref{fig:mrCorr}) comparing the 2D RPA effective mass with the 2D
effective mass calculated in the Hubbard approximation. The basic idea
is to introduce an approximate vertex correction (to both the
polarizability function and the self-energy) in the theory to
calculate the renormalized effective mass in order to assess the
robustness of the RPA approximation. A rigorous calculation of the
self-energy vertex correction in the dynamical screening expansion is
essentially an impossibly formidable task; one must therefore resort
to drastic approximations. One reasonably successful technique for
introducing vertex corrections to the theory has been to preserve the
formal structure of the RPA self-energy and the polarizability with a
new function $\gamma$, the so-called local field correction, modifying
the electron-electron interaction mimicking the higher-order vertex
corrections. Formally, the self-energy is given by Eqs.~(\ref{eq:EV})
and (\ref{eq:epsilonV}), and the $\gamma$ function is given in terms
of a local field factor $G(k)$ by Eqs.~(\ref{eq:gamma}) and
(\ref{eq:Gfactor}). $G(k)$ can be chosen satisfying various sum rules.
The two most popular forms for the local field correction are due to
Hubbard~\cite{hubbard}, and in 2D system they take the form
\begin{eqnarray}
\label{eq:HA}
\mbox{HA1}: G(k) &=& {k \over 2 \sqrt{k^2 + k_F^2}}, \\
\mbox{HA2}: G(k) &=& {k \over 2 \sqrt{k^2 + k_F^2 + q_{TF}^2}}. 
\end{eqnarray}
We have already used the first model (HA1) in obtaining results in
Fig.~\ref{fig:mrCorr}. We find a divergent $m^*(r_s)$ for $r_s \sim 10
- 17$ in all three approximations with HA1 giving a stronger
divergence than the other two. In Fig.~\ref{fig:mrCorr2} we compare
these two vertex correction approximations with RPA for the calculated
2D $m^*(r_s)$ as a function of $r_s$ in the $r_s = 0 - 17$ regime. The
results show that (i) qualitatively all three approximations give
similar results, and (ii) the HA2, which is considered~\cite{rice} to
be a superior vertex correction than HA1, actually produces $m^*(r_s)$
quantitatively very close to RPA.

The results of Fig.~\ref{fig:mrCorr2}, which include vertex
corrections, again confirm the essential qualitative validity of our
calculated $m^*(r_s)$. The point we are making here is that our main
qualitative conclusion that $m^*(r_s)$ diverges at some value of $r_s$
remains valid even when Hubbard approximation the vertex correction is
made by going beyond RPA. On the other hand the precise value of $r_s$
where $m^*$ diverges may depend on the approximation.

It is well-known that the strongly interacting Coulomb system at large
$r_s$, where the electron-electron interaction cannot be considered a
weak perturbation to the noninteracting kinetic energy, does not allow
for a controlled systematic many-body calculation since no
order-by-order perturbative diagrammatic expansion is possible because
there is no small parameter (in contrast to the high-density
situation, $r_s \ll 1$, where the interaction is a weak perturbation
and the RPA self-energy expansion is asymptotically exact). The great
merit of RPA lies in the fact that it is a systematic self-consistent
mean-field theory (apart from also being an exact low-$r_s$
perturbative theory) because it includes the basic physics of the
dynamical screening of the long-range Coulomb interaction.  As such
RPA should remain qualitatively valid at large $r_s$ until there is a
quantum phase transition to a new state.  There have been many
systematic attempts~\cite{add} to go beyond RPA in the literature in
calculating the electron self-energy, and many of these attempts
include the approximate vertex corrections by exploiting the
self-consistent field (rather than the perturbative) nature of RPA by
including the vertex corrections through various local field factors.
The two Hubbard approximations we include in our work are only among
the simplest such approximations incorporating vertex corrections-- we
refer to the literature~\cite{add} for more sophisticated treatments
of vertex corrections. The important point in the context of our
theory is that the incorporation of vertex corrections is unlikely to
cause a qualitative change in the RPA results although quantitative
changes are expected. Therefore, we can conclude that our finding of
an effective mass divergence within the RPA self-energy calculation
(and the Hubbard approximation) should remain valid in general
although the critical $r_s$ where the mass divergence occurs is likely
to be affected by the specific many-body approximation.


\section{Discussion and Conclusion}
\label{sec:con}

We have presented in this paper a detailed quantitative calculation
for the zero-temperature quasiparticle effective mass in a strongly
interacting homogeneous electron liquid using the realistic long-range
Coulomb interaction and the noninteracting parabolic kinetic energy
dispersion. The theory covers a large range of the dimensionless
interaction parameter space, and provides $m^*(r_s)$ down to the
low-density regime $r_s \sim 15 - 20$.  We have focused primarily on
the (semiconductor-based) 2D electron systems since rather high (low)
values of $r_s$ (density) are now routinely achieved experimentally in
2D semiconductor systems, and therefore there is considerable current
interest in the many-body properties of 2DES. Our main results for
$m^*(r_s)$ are qualitatively similar in 2D and 3D systems. We also
provide in this paper a serious and substantive theoretical analysis
of the feasible many-body approximations for strongly interacting
electron liquids within the Feynman-Dyson diagrammatic perturbation
theory. We argue (based on a number of approximate non-rigorous
arguments) that the RPA self-energy expansion in the dynamically
screened Coulomb interaction, the dynamically screened single-loop
Hatree-Fock (or equivalently `$G_0 u$' or `$G W$') approximation, may
very well be a reasonably good (albeit somewhat uncontrolled in the
sense that a systematic improvement on the single-loop `$G_0 u$'
approximation is difficult) approximation even at large $r_s$ (while
being exact in the $r_s \to 0$ limit, as has been known for a long
time). In the process we establish that the RPA self-energy or
effective mass calculation is {\em not} an $r_s$ expansion at
arbitrary $r_s$ values, it is more likely to be an expansion in an
effective parameter $\delta = (r_s /c) /(a + r_s)$, where $a \sim 1$
and $c > 1$ in both 2D and 3D. Thus, RPA is exact in the $r_s \to 0$
limit, and is reasonably well-valid for $r_s \gg 1$ provided $c \gg
1$; we find $c \approx 4 \pi$ (2D), $6 \pi$ (3D) showing that the RPA
self-energy and effective mass calculation may have reasonable
qualitative (perhaps even quantitative) validity in the large $r_s$
regime of our interest (provided that the strong interaction does {\em
  not} drive the system to a quantum phase transition). The precise
level of the quantitative validity of RPA cannot, however, be
ascertained at large $r_s$ values since we have no reliable way of
estimating the terms being left out of the calculation. (In this
respect RPA shares the shortcomings of other well-known electronic
theory approximations such as LDA or DMFT which are very successful
quantitative theories without being controlled systematic
approximations.)  Analyzing in some depth the structure of the
infinite resummation of perturbation terms involved in the RPA
expansion in the screened interaction for the self-energy function, we
also establish rather compellingly that the on-shell RPA quasiparticle
energy (and effective mass) calculation is the appropriate correct
approximation at arbitrary $r_s$ values, and the off-shell
approximation in fact leads to qualitatively incorrect results (by
virtue of the inconsistent mixing of different orders) at large $r_s$;
for small $r_s ( \ll 1)$, of course, the on-shell and the off-shell
approximations become equivalent.

We emphasize that, in spite of several crude (and at best persuasive,
by no means rigorous and compelling) arguments we have provided in
section~\ref{sec:RPA} in support of our approximation scheme, our
results can only be considered qualitative since the ring-diagram
approximation inherent in RPA is exactly only in the high density
($r_s \to 0$) limit whereas our finding of a low-density effective
mass divergence occurs in the strongly interacting $r_s \approx 10-15$
regime. While agreeing that RPA is, in principle, an uncontrolled
many-body approximation at large $r_s$, we claim that there is no
particular value of $r_s$ at (or above) which the RPA self-energy
fails {\it qualitatively}. In fact, the evidence is quite contrary,
typically RPA predictions for qualitative behavior {\em always} turn
out to be valid. For example, RPA energetics for interacting electron
systems predict a ferromagnetic instability at $r_s \approx 19(7)$ in
3D (2D) electron systems where the spin $g$-factor (or equivalently,
the static susceptibility) diverges, and it is now universally
accepted that an interacting electron system indeed undergoes an
interaction-driven low-density ferromagnetic instability ( the
so-called Bloch ferromagnetism) albeit at $r_s$ values larger than the
RPA predictions. We see no reason why our finding of a low-density
effective mass divergence within RPA should be any different from the
corresponding divergence in the magnetic susceptibility. It is likely
that RPA underestimates the critical $r_s$ where the effective mass
diverges, and the real effective mass divergence would occur at higher
$r_s$ values, but we see no reason for RPA to give a qualitatively
incorrect answer for the effective mass divergence. An important
related point in this context is that the RPA or the ``GW
approximation'' is the {\em only} available many-body approximation
for quantitative electronic self-energy calculations, and as such it
is routinely used for 3D metallic systems with $r_s \approx 5$. In 2D
electron systems, RPA effective mass calculation has earlier been
carried out in the literature up to $r_s \approx 5$ also in the
context of providing a quantitative explanation for effective mass
measurements in 2D semiconductor structure. We have now extended these
calculations up to $r_s \approx 10-15$ because recent low density 2D
systems have achieved such large values by providing high-quality
low-density 2D structures.

Our theory involves two distinct approximations: RPA or ring-diagram
approximation for the electronic self-energy calculation and the
on-shell approximation for calculating the quasiparticle energy
leading to the quasiparticle effective mass. While the RPA self-energy
calculation is an essential (and in some sense, uncontrolled)
approximation in our theory (except for incorporating vertex
corrections through crude local field approximations, e.g. the Hubbard
approximation, as we do in some of our calculations), one could
question the need (or even the validity) of the on-shell quasiparticle
effective calculation. We have argued in this paper that for our `$G_0
u$' self-energy approximation (Fig.~\ref{fig:feynman}), the on-shell
effective mass calculation is, in fact, the appropriate approximation
(rather than the off-shell approximation which solves the full Dyson's
equation iteratively). We assert that, as shown by Rice~\cite{rice}
and used by Quinn and collaborators~\cite{ting} a long time ago, the
effective mass approximation consistent with the Landau Fermi liquid
theory and our `$G_0 u$' self-energy approximation is precisely the
on-shell approximation. This is simply because we use $G_0$, the
non-interacting Green's function in all our internal propagator, and
therefore the quasiparticle energy consistent with this leading-order
self-energy calculation is precisely our Eq.~(\ref{eq:onshellE}), the
on-shell approximation, rather than our Eq.~(\ref{eq:offshellE}), the
off-shell approximation. This has been discussed in great details by
Rice~\cite{rice} almost forty years ago (and was already implemented
for 2D systems in Ref.~\cite{ting} almost thirty years ago). Our
on-shell quasiparticle effective mass is therefore precisely the Fermi
liquid quasiparticle effective mass~\cite{yakovenko}.

We find that the quasiparticle effective mass $m^*(r_s)$ at $T=0$
increases rapidly (both in 2D and 3D) at low densities with increasing
$r_s$, consistent with the strongly interacting nature of the Coulomb
system at large values of $r_s$. One most important and interesting
result is a true divergence of $m^*(r_s)$ at a critical $r_s \sim
16.6$ in 2D (the mass divergence occurs around $r_s \sim 48$ in 3D)
within the on-shell single-loop $G_0 u$ RPA self-energy calculation.
This theoretical effective mass divergence (or at least the rapid
increase of effective mass with decreasing density) seems to
superficially simulate recent measurements~\cite{kravchenko1,
  kravchenko2, pudalov} of low density 2D effective mass (from the
temperature dependence of the SdH oscillations of the low field
magnetoresistance) in n-Si MOSFETs.

There are three general questions that immediately arise in the
context of our theoretical finding of a divergent $m^*(r_s)$ for
sufficiently large (but experimentally accessible in 2D, but {\em not}
in 3D systems) values of the density parameter $r_s$. These questions
are: (i) Are our theoretical results of any relevance to experiments
(particularly to the large body of 2D M-I-T literature where
large-$r_s$ behavior of 2DES is being explored extensively)? (ii) Is
our finding of a divergent effective mass at a large critical value of
$r_s$ real? (i.e. is it a truly robust finding as we have argued in
this work or just an artifact of our single-loop $G_0 u$ approximation
scheme?) (iii) If the interaction-driven divergence of $m^*(r_s)$ is
real, then what does this phenomenon imply for an electron liquid?
Below we discuss some partial answers to (and educated speculations
on) these questions. Since an exact solution to this strongly
interacting (and 50-year old) many-body problem is unlikely to come in
the near future, we can only discuss various aspects of this problem
qualitatively with no definitive resolution of these important
questions.

\begin{figure}[htbp]
\centering \includegraphics[width=3in]{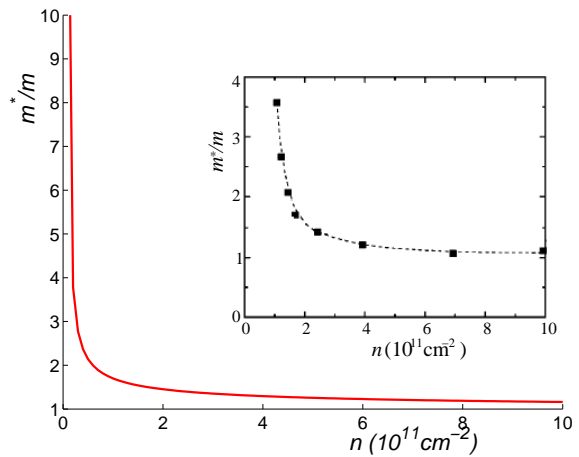}
  \caption{Calculated RPA effective mass within the on-shell 
    approximation at different density values for 2D Si MOSFET system.
    Inset: Experimental effective mass results with data taken from
    Ref.~\cite{kravchenko1}.}
\label{fig:real}
\end{figure}

For whatever it is worth we can, of course, carry out a direct
comparison with the experimentally measured density-dependent
effective mass with our calculated $m^*(r_s) \equiv m^*(n)$ for the
appropriate 2DES. We show our theoretical results for the $T=0$ RPA
effective mass as a function of the carrier density in Si-SiO$_2$
inversion layer in Fig.~\ref{fig:real}, including the quasi-2D
form-factor effects in the Coulomb interaction by modifying the bare
Coulomb interaction $v_q = 2 \pi e^2/({\bar \kappa} q)$ to $v(q) F(q
b)$ where ${\bar \kappa}$ is the effective background dielectric
constant for the system, `$F$' is the quasi-2D form-factor, and $b
\equiv b(n)$ is the finite quasi-2D layer width parameter discussed in
section~\ref{sec:form:W}. For Fig.~\ref{fig:real} results we have used
the appropriate semiconductor electron band mass and lattice
dielectric constant (given in Eqs.~(\ref{eq:cSi})) in the calculation.
In Fig.~\ref{fig:real} we have shown as an inset the experimental Si
MOS effective mass values for a comparison with our theory. Clearly,
the theory and experiment are in {\em excellent} qualitative
agreement, most notably in the rapidly rising low density $m^*(n)$
behavior. While it is certainly gratifying to see that our theory
agrees well with the existing low-density effective mass measurements,
one must take such agreements with more than a grain of salt. The
reliability and/or the level of accuracy of the experimentally
measured $m^*(n)$, particularly at the low carrier densities of
interest to us, is unknown. This is particularly true in view of the
experimental effective mass measurements being based on the
temperature-dependent amplitude of the weak-field SdH
magnetoresistance oscillations at each density. We have shown
elsewhere~\cite{short} that the quasiparticle effective mass
$m^*(r_s)$ has a strong temperature dependence at low densities -- in
fact $m^*(T/T_F) \sim m^*(0) [1 + c (T/T_F)]$ with $c>0$. Such an
intrinsic temperature dependence of the effective mass itself may
introduce unknown errors in the experimental mass measurements. In
addition, there is always the key problem of not quite knowing what
the electron temperature really is (a particularly severe problem at
low carrier densities) which could lead to larger errors in the
experimental determination of $m^*(n)$. In spite of these caveats the
excellent qualitative agreement between theory and experiment
demonstrated in Fig.~\ref{fig:real} is quite impressive.

In addressing the issue of the ``robustness'' of our finding of a
divergence of $m^*(r_s)$ at a large critical $r_s$ we first assert
that within our approximation scheme this interesting result is
certainly real and not an artifact as demonstrated by the manifest
model independence of our result -- we find that $m^*(r_s)$ diverges
at a large $r_s$ both in 2D and 3D systems, both within the simple RPA
and the Hubbard approximation calculations, and both in the strictly
2D and quasi-2D systems. The only model dependence of our theory is in
our finding that the off-shell approximation (i.e. the full solution
of the Dyson's integral equation using, however, only the leading
order dynamically screened single-loop self-energy function) does {\em
  not} leads to a mass divergence (and in fact lead instead to an
effective mass saturation at large $r_s$ in clear disagreement with
experimental results), but we have given convincing arguments that the
off-shell approximation is theoretically unjustified ( and
inconsistent) in view of the single-loop leading-order nature of our
`$G_0 u$' self-energy calculation.

Although the divergence of $m^*(r_s)$ at large $r_s$ is certainly a
real result within our `$G_0 u$' single-loop self-energy approximation
scheme, the important question of whether it is a correct result (i.e.
valid beyond our approximation scheme) for the interacting electron
liquid remains unanswered at this stage. While we have included some
approximate vertex corrections (see section~\ref{sec:RPA}) in the
theory, a self-energy calculation going beyond the leading-order `$G_0
u$' single-loop diagram is essentially impossible. The second-order
self-energy diagrams in the dynamically screened Coulomb interaction
expansion (Fig.~\ref{fig:feynman2}) involve six-dimensional
highly-singular (with overlapping singularities and branch cuts)
integrals in the `$G_0 u G_0 G_0 u$' second-order terms. We see no
hope of an accurate (numerical) evaluation of these 6-dimensional
highly singular principal value integrals in order to reliably
calculate $\Sigma({\bf k}, \omega)$ to the 2-loop level so that
$m^*(r_s)$ can be calculated correctly (formally) to second order in
the dynamically screened interaction. Such a (hopeless) calculation is
necessary to see whether the divergence of $m^*(r_s)$ is truly a
feature of the interacting electron systems at low carrier densities
or is merely an artifact of the single-loop self-energy formalism.
Even such a 2-loop calculation will {\em not} obviously definitively
establish the existence of a density-driven effective mass divergence
transition (because one could still wonder what would happen at the
3-loop or the 4-loop level), but we believe that finding a divergence
of $m^*(r_s)$ in both 1-loop (as we do here) and 2-loop level will go
a long way in establishing the very strong possibility that such a
divergence does indeed occur in the ground state of an interacting
quantum Coulomb Fermi system.

Short of such a 2-loop self-energy calculation (which we believe to be
impossibly difficult) we have to make do with a number of heuristic
arguments supporting the reality of the effective mass divergence
reported in this work. First, the single-loop $G_0 u$ RPA self-energy
calculation (along with its simple Hubbard approximation vertex
correction) certainly leads to an effective mass divergence. Second,
we give several arguments (see section~\ref{sec:RPA}) why we believe
the $G_0 u$ approximation to be at least qualitatively valid at large
$r_s$ where the effective mass divergence occurs. Third, the structure
of the effective mass formula (see section~\ref{sec:form}) indicates
that the necessary and the sufficient condition for the divergence of
$m^*(r_s)$ is to have ${d \Sigma \over d k} = -1$, which is certainly
allowed as a matter of principle. Fourth, the fact that we get the
effective mass divergence in both 2D and 3D systems indicates that the
effective mass divergence may be a generic feature of interacting
electron systems in the strong interaction regime. Fifth, as mentioned
in the Introduction, the strongly interacting normal He-3 (with strong
short-range interactions) is also thought to be an almost-localized
Fermi liquid (i.e. very large quasiparticle effective mass) akin to
what we find for strongly interacting electron liquids interacting via
the Coulomb interaction. Similar to our finding in this work, normal
He-3 is also reported~\cite{casey, boronat, vollhardt} to go through
an effective mass divergence transition at sufficiently large
interaction strength.  It is therefore quite possible, perhaps even
very likely, that both short-range interacting He-3 and long-range
interacting electron liquids have quasiparticle effective mass
divergence at sufficiently large interaction strength; such an
effective mass divergence leading to ``an almost-localized Fermi
liquid'' may very well be a generic feature of strongly interacting
quantum Fermi systems.

Accepting that there is indeed a critical (large) value of $r_s \equiv
r_c$ ($\approx 16.6$ in the ideal 2DES according to the single-loop
calculation, but in general we expect $r_c$ to be some large $r_s$
value indicating the strong interaction regime) where $\left. m^*(r_s)
\right|_{r_s \to r_c}$ diverges, the important question of the nature
and implications of this mass divergence takes on great significance
(both theoretical and practical since 2DES with $r_s > 10$ is
routinely available with the state of the arts semiconductor
technology). We first note that according to the Landau's Fermi liquid
theory the low-temperature thermodynamic properties such as specific
heat, spin susceptibility, and compressibility are all proportional to
the quasiparticle effective mass:
\begin{eqnarray}
\label{eq:FermiPara}
{C_v^* \over C_v} &=& {m^* \over m} = 1 + A_1, \nonumber \\
{\chi^* \over \chi} &=& {m^* g^* \over m g} 
= {m^* \over m} {1 \over 1 + B_0}, \nonumber \\
{\kappa^* \over \kappa} &=& {m^* \over m} {1 \over 1 + A_0},
\end{eqnarray}
where $A_0$, $B_0$, $A_1$ are the appropriate Landau Fermi liquid
parameters. Assuming that $A_0$ and $B_0$ are well-behaved at low
densities, we conclude that both the spin susceptibility and the
compressibility will also diverge around the critical $r_s$ where the
effective mass diverges. Such anomalous behavior in the magnetic
susceptibility, with rapidly increasing susceptibility at low
densities, has indeed been reported in several recent experimental
studies~\cite{kravchenko3, vitkalov} of 2DES. Recent 2D experiments
also report~\cite{dultz} the inverse compressibility passing through
zero at low carrier densities indicating an infinite compressibility
consistent with the mass divergence (assuming regular density
dependent behavior of the Landau parameter $1 + A_0$). But the
compressibility measurements~\cite{dultz} are most likely connected
with disorder effects invariably present in the system, and are
unlikely to be intrinsic Fermi liquid renormalization phenomena. More
work is needed to settle these issues.

A key question in this context (accepting the mass divergence to be
real) is whether this effective mass divergence signifies a new strong
coupling fixed point of the interacting electron system or is a subtle
manifestation of one of the already known quantum phase transitions
which may occur in an interacting electron system. A related question
of equal importance is whether this mass divergence necessarily
implies a breakdown of the Fermi liquid theory. Unfortunately, none of
these questions can be answered at our current stage of understanding
of the subject since our calculation only leads to the effective mass
divergence without telling us the nature of the new ground state, if
indeed there is a quantum phase transition associated with the mass
divergence phenomenon. Using understanding~\cite{vollhardt} developed
in the context of the strongly interacting neutral He-3, which also
undergoes~\cite{casey, boronat} a very similar effective mass
divergence transition, we speculate that the transition associated
with the effective mass divergence may very well be a continuum
version of a Mott transition, where increasing Coulomb interaction in
a system leads to ``localization'' and metal-insulator transition as
the system can lower its interaction energy at the ``cost'' of kinetic
energy by developing infinite mass (i.e. ``no hopping'') -- a
divergent mass in a continuum system is the closest analog to the
band-width collapse due to the disappearance of hopping in the Mott
transition in a lattice system. It is generally assumed that the
Wigner crystallization transition, which in 2D (3D) systems is
thought~\cite{tanatar, ceperley} to occur at large $r_s$ values $\sim
38 (72)$, is the continuum analog of the Mott transition, but it is,
in principle, possible for the system to undergo a different type of
localization transition through the effective mass divergence long
before the Wigner crystallization low-density point is reached. The
alternate possibility is that our effective mass divergence is
essentially a signature or a precursor to the eventual Wigner
solidification transition in the system. We also can not rule out the
possibility that the RPA effective mass divergence is a precursor of a
charge density wave transition in the system (related to, but not
identical to, the Wigner transition). Any of these transitions of
course signifies a breakdown of the normal Fermi liquid picture
because the one-to-one correspondence with the noninteracting electron
gas ground state is destroyed beyond the transition point. At this
stage, however, we do not rigorously know the nature of the
interaction-driven effective mass divergence phenomenon (except that
it happens at some large $r_s$ values). Much more work will be needed
to develop an understanding of the mass divergence phenomenon reported
in this work, but the very real exciting possibility exists that all
strongly interacting homogeneous Fermi liquids undergo a continuum
version of a Mott transition, where the renormalized quasiparticle
effective mass diverges, at sufficiently high interaction strengths.
At this stage we do not know whether our theoretical discovery of a
low-density divergence in the quasiparticle effective mass liquefies
any kind of phase transition or not. In may be worthwhile to point out
in this context that the corresponding problem of a {\em single}
electron strong coupled to the lattice (``the strong coupling
polaron'' problem) also shows a sharp effective polaronic mass
divergence at large electron-phonon coupling strength~\cite{polaron},
which is superficially very similar to the quasiparticle effective
mass divergence we find in the current interacting electron problem. A
key difference between the polaronic problem and current interacting
electron problem is the fact that the polaronic case deals with a {\em
  single} electron strongly coupled to a bosonic bath (i.e. phonons)
whereas the current problem involves infinite fermionic degrees of
freedom (i.e. this is a many-electron problem). Thus, the polaronic
effective mass divergence (the so-called self-trapping of the
polaron), being a one electron problem, cannot by definition involve
any phase transition whereas the current problem being a problem with
infinite degrees of freedom does (at least) allow the possibility of a
phase transition. Whether the quasiparticle effective mass divergence
reported in this work is indeed a phase transition or not remains to
established through further work.

This work is supported by the ONR, the LPS, and the NSF.

Six months after the submission of our manuscript for publication and
its posting on the archive (cond-mat/0312565), a recent preprint
(cond-mat/0406676) has appeared which verifies our finding of the
quasiparticle effective mass divergence in low density interacting
electron liquids at a critical density.  In this
preprint(cond-mat/0406676), the quasiparticle effective mass
divergence is found to occur even within many-body local field
approximations which are more sophisticated than the simple RPA and
Hubbard approximation many-body theories utilized in our work.  The
theoretical finding of a quasipasrticle effective mass divergence in
interacting electron liquids by two different groups employing
different approximation schemes give us confidence that the mass
divergence is a true many-body phenomenon, and not an artifact of a
specific approximation scheme.


\bibliography{diverge}

\end{document}